\documentclass[final,5p,times,twocolumn]{elsarticle}
\usepackage{graphicx}
\usepackage{natbib}
\usepackage{amsmath}
\usepackage{amssymb}
\usepackage{amsfonts}
\usepackage{mathtools}
\usepackage[super]{nth}
\usepackage{setspace}
\usepackage{enumitem} 
\usepackage{epstopdf}
\usepackage{color}
\usepackage{algpseudocode}
\usepackage{algorithm}
\usepackage{multirow}
\usepackage{longtable}
\usepackage{rotating}
\usepackage{mathrsfs}
\usepackage{titlesec}
\usepackage{subfigure}
\usepackage{subcaption}
\usepackage{float}
\usepackage{graphicx}
\usepackage{booktabs}
\usepackage{caption}
\usepackage{bm}
\usepackage{tikz}
\usepackage{makecell}
\usepackage[toc,title,page]{appendix}
\newtheorem{remark}{Remark}
\titleformat{\paragraph}
{\normalfont\normalsize\bfseries}{\theparagraph}{1em}{}
\journal{Control Engineering Practice}

\begin{document}
\begin{frontmatter}
\title{Sensitivity-Informed Parameter Selection for Improved Soil Moisture Estimation from Remote Sensing Data}
\author[label1]{Bernard T Agyeman}
\author[label1]{Erfan Orouskhani }
\author[label2]{Mohamed Naouri}
\author[label2]{Willemijn M. Appels}
\author[label3]{Maik Wolleben}
\author[label1]{Jinfeng Liu}
\author[label1]{Sirish L. Shah}

\affiliation[label1]{organization={Department of Chemical \& Materials Engineering, University of Alberta},
            city={Edmonton},
            postcode={T6G 1H9},
            state={Alberta, AB},
            country={Canada}}

\affiliation[label2]{organization={Centre for Applied Research, Innovation and Entrepreneurship, Lethbridge Polytechnic},
            city={Lethbridge},
            postcode={T1K 1L6},
            state={Alberta, AB},
	country={Canada}}

\affiliation[label3]{organization={Skaha Remote Sensing Ltd.},
	city={Surrey},
	postcode={V3W 1K7},
	state={British Columbia, BC},
	country={Canada}}
	
\begin{abstract}
Improving the accuracy of soil moisture estimation is required for advancing irrigation scheduling and water conservation efforts. Central to this task are soil hydraulic parameters, which govern moisture dynamics but are rarely known precisely and must therefore be inferred from observational data. In large-scale agricultural fields, estimating the complete set of these parameters is often impractical due to the sparse and noisy nature of available measurements. To address this challenge, this work develops a framework that uses sensitivity analysis and orthogonal projection to identify parameters that are both reliably estimable from available data. These parameters, together with the spatial distribution of soil moisture, are jointly estimated by assimilating observational data into a cylindrical-coordinate version of the Richards equation using an extended Kalman filter. The soil moisture measurements are obtained from microwave remote sensors mounted on center pivot irrigation systems---an emerging and practical technology for capturing field-scale variability. Numerical simulations and field experiments conducted on a large-scale site in Lethbridge, Alberta, Canada, demonstrate that the proposed method improves soil moisture estimation accuracy by 24–43\% and enhances predictive model performance by 50\%. Furthermore, the estimated parameters—particularly saturated hydraulic conductivity—exhibit good agreement with experimental measurements.	
\end{abstract}


\end{frontmatter}

\section{Introduction}
Agriculture accounts for nearly 70\% of all freshwater withdrawals, with irrigation activities responsible for the largest share~\cite{unwater2015}. As freshwater resources become increasingly scarce, it is essential to adopt strategies that address the worsening supply crisis. Improving irrigation water-use efficiency offers an effective response, given agriculture’s significant share of global water demand. One such approach is closed-loop irrigation, which enhances efficiency by using real-time feedback to guide water application~\cite{SBL21AIChE}.

Soil moisture sensing is essential for feedback control in closed-loop irrigation systems. Non-invasive methods, such as microwave remote sensing, offer a practical solution for monitoring moisture distribution across large-scale agricultural fields. Microwave sensors are capable of quantitatively measuring near-surface soil moisture under varying field conditions. A growing application involves installing these sensors on center pivot irrigation systems---mechanized systems that irrigate crops in a circular pattern around a central pivot. In this configuration, the sensors collect spatially distributed measurements as the pivot rotates, and soil moisture information is typically aggregated into moisture maps at the end of each rotation cycle.

Soil moisture sensing techniques are often unable to provide continuous and spatially comprehensive observations, resulting in spatio-temporal data gaps. Relying solely on such incomplete information to operate closed-loop irrigation systems can lead to inefficient water use. A practical solution to this limitation is sequential data assimilation—also known as state estimation—which combines sparse sensor measurements with a dynamic model to estimate the full state of the system. By leveraging model dynamics, sequential data assimilation enables interpolation and extrapolation of observations, resulting in a more complete and consistent estimate of soil moisture distribution.
Several techniques have been applied in this context, including the extended Kalman filter (EKF)\cite{lu2011dual,sabater2007near}, the ensemble Kalman filter (EnKF)~\cite{Medina2014,reichle2002hydrologic}, the particle filter~\cite{montzka2011hydraulic,pan2008estimation}, and the moving horizon estimator (MHE)\cite{bo2020parameter,bo2020decentralized}. For instance,~\cite{reichle2002hydrologic} used the EnKF to estimate soil moisture by assimilating L-band microwave radiobrightness observations into a land surface model. In~\cite{bo2020parameter}, the MHE was used to jointly estimate soil moisture and hydraulic parameters using the 1D Richards equation. More recently,~\cite{agyeman2021soil} introduced an information fusion system that integrates the cylindrical-coordinate form of the Richards equation, the EKF, and microwave remote sensing measurements to estimate soil moisture in fields equipped with center pivots.

Accurate simulation of soil moisture in agro-hydrological models requires well-defined soil hydraulic parameters. These parameters are traditionally estimated through laboratory measurements or inferred using pedo-transfer functions (PTFs). However, laboratory methods are often labor-intensive and impractical for large-scale applications, while PTFs can yield uncertain results in heterogeneous field conditions~\cite{li2007estimating,soet2003functional}. Moreover, both approaches provide static estimates that cannot capture temporal variability in soil properties. A practical alternative is to retrieve soil hydraulic parameters directly from real-time soil moisture observations. This approach allows for dynamic estimation of both soil moisture states and model parameters within a unified framework.

Given the coupled nature of soil moisture dynamics and hydraulic properties in agro-hydrological systems, a simultaneous estimation approach is particularly well suited. By treating soil hydraulic parameters as augmented states, this method captures their interdependence with soil moisture and enables more accurate estimation within agro-hydrological modeling frameworks. This joint estimation strategy has been widely explored in the literature. For instance,~\cite{li2011estimation} employed EnKF to estimate soil moisture and hydraulic parameters simultaneously. Medina et al.~\cite{medina2014kalman} proposed a hybrid scheme combining Kalman and unscented Kalman filters for joint state and parameter estimation. 

Parameter identifiability is an important consideration in the estimation of dynamic models, as it determines whether model parameters can be uniquely inferred from available observations. Assessing  parameter identifiability is essential for at least one primary reason: regardless of the data assimilation technique used, non-identifiable parameters can compromise the reliability of parameter estimates~\cite{chis2011structural}. In much of the soil moisture and hydraulic parameter estimation literature, identifiability is often evaluated only after estimation---typically using the covariance matrix of the estimated parameters. While this matrix can reveal parameter dependencies through high off-diagonal correlations, it has significant limitations, especially in multi-parameter systems. As noted in~\cite{kuczera1990assessing}, interpreting the covariance structure becomes increasingly unreliable as the number of estimated parameters increases. Moreover, post-estimation analysis often results in wasted computational effort, as non-identifiable parameters are included in the estimation process despite contributing little or no meaningful information. To address the limitations associated with post-estimation indentifiablity assessments, several studies in the general context of state and parameter estimation have advocated for assessing parameter identifiability prior to model calibration~\cite{saccomani2003parameter,yao2003modeling}. Pre-estimation analysis ensures that only structurally identifiable parameters are targeted, thereby improving both the accuracy and efficiency of the estimation process.

Sensitivity analysis is a widely used technique for evaluating parameter identifiability, as it quantifies how variations in parameter values affect model outputs~\cite{walter2013identifiability}. This is typically carried out by computing a sensitivity matrix at nominal values of the parameters. When this matrix has full column rank, it indicates that all parameters are structurally identifiable and can, in principle, be uniquely estimated from the available observations. However, in large-scale systems (e.g., agro-hydrological models applied to field-scale environments), this condition is rarely met. These models often include a high number of parameters, while available measurements are limited in both quantity and precision. As a result, the sensitivity matrix tends to be rank-deficient, reflecting the fact that only a subset of parameters can be reliably estimated~\cite{yao2003modeling,liu2021simultaneous}.

A practical workaround in the face of a rank-deficient sensitivity matrix is to apply parameter selection methods that identify a subset of parameters which can be estimated given the available data. Techniques such as principal component analysis (PCA) and the orthogonal projection method are commonly used for this purpose~\cite{kravaris2013advances}. While PCA reduces dimensionality by forming new variables from linear combinations of parameters, it often sacrifices physical interpretability. In contrast, the orthogonal projection method identifies a subset of the original model parameters that are both estimable and physically meaningful. This makes it particularly well suited for physically based models, where maintaining interpretability is essential.

The literature on estimation of dynamical systems has extensively explored the use of parameter identifiability analysis and parameter selection to improve the reliability and accuracy of parameter estimates, and to ensure that the estimation problem is well-posed. Sensitivity analysis and orthogonal projection have been key tools in this regard, with successful applications across various domains. For example, in mechanistic models of ethylene copolymerization, these techniques have been used prior to estimation to retain only structurally identifiable parameters~\cite{yao2003modeling}. In contrast, such methods have seen limited application in the estimation of soil moisture and hydraulic parameters in agro-hydrological systems. To the best of current knowledge, the study by~\cite{bo2020parameter} is among the first to conduct an identifiability assessment prior to estimation. Their results showed that incorporating identifiability analysis and parameter selection improved the  accuracy of the estimated soil moisture and hydraulic properties. However, the parameter selection procedure relied on a simple heuristic---ranking and selecting parameters based on the magnitudes of the sensitivity matrix column norms. While computationally straightforward, this approach overlooks linear dependencies among sensitivity vectors, which can affect identifiability even when column magnitudes are large.

Beyond the limited adoption of pre-estimation parameter identifiability and selection analyses in agro-hydrological models, applying these techniques to large-scale fields introduces significant challenges. As field size increases, so does the heterogeneity of soil properties, often necessitating a greater number of localized parameters to accurately capture hydraulic behavior. This rise in parameter dimensionality increases the computational burden of identifiability assessments and complicates the task of determining a subset of reliably estimable parameters. In this study, these challenges are further intensified by the use of remotely sensed soil moisture data collected from sensors mounted on center pivot irrigation systems. Unlike fixed-location measurements, these observations vary across both space and time. This spatial variability complicates the construction and analysis of the sensitivity matrix, impacts the parameter selection process, and introduces additional complexity into the data assimilation procedure. Collectively, these challenges point to the need for frameworks that unify parameter identifiability analysis, selection, and data assimilation---capable of addressing the scale, heterogeneity, and spatio-temporal complexity of observational data in modern agro-hydrological systems.

This study aims to improve the accuracy and reliability of soil moisture estimation in large-scale agro-hydrological systems by systematically addressing challenges related to parameter identifiability. The proposed approach involves identifying a subset of soil hydraulic parameters that are both structurally identifiable and physically meaningful, and simultaneously estimating these parameters alongside soil moisture states using an extended Kalman filter applied to a cylindrical-coordinate formulation of the Richards equation. Soil moisture observations are obtained from microwave sensors mounted on a center pivot irrigation system. To enhance spatial consistency, Kriging interpolation is used to extend parameter estimates across the entire field.

The study is guided by the hypothesis that performing parameter identifiability analysis in advance—followed by the selection of a subset of estimable parameters when the full set cannot be reliably inferred—enhances the overall accuracy of soil moisture estimates and produces parameter values that are physically interpretable. This hypothesis is evaluated through a combination of simulation studies and field experiments conducted on a large-scale agricultural field.

The key contributions of this work are summarized as follows:
\begin{enumerate}
	\item A procedure for constructing and evaluating the sensitivity matrix in systems with spatially varying measurements. In addition, a method is developed to perform parameter selection using the orthogonal projection technique adapted for mobile sensing scenarios.
	\item A computationally efficient modification of the extended Kalman filter that accommodates time-varying sets of estimable parameters.
	\item A comprehensive evaluation of the proposed framework through simulations and field experiments, demonstrating improved accuracy in both state and parameter estimation.
\end{enumerate}

This study builds upon preliminary findings reported in~\cite{orouskhani2022simultaneous,agyeman2022simultaneous}, but extends them in several key ways. Unlike the earlier work, the present study offers more comprehensive theoretical explanations and includes a broader set of simulation experiments designed to evaluate the influence of sensitivity-based parameter selection on estimation performance. In addition, a wider range of validation scenarios is incorporated in the field case study to  assess the practical effectiveness of the proposed methodology.

\section{Model Development}
\label{sec:System Description}
\subsection{Field model}
The system under study in this work is an agro-hydrological system equipped with a center pivot irrigation system as illustrated in Figure~\ref{fig:Polar_Agrohydrological}. In this work, the 3D version of the Richards equation, in cylindrical coordinates, is selected as the field model. This choice of the 3D model is motivated by two key factors. Firstly, considering that the microwave radiometers measure soil water content as the center pivot rotates and irrigates the field under study, it is crucial to incorporate this rotation into the field model. This ensures accurate depiction of the irrigated areas at each time step as well as an accurate depiction of the specific locations being measured at each time step. The cylindrical coordinate-based 3D version of the Richards equation naturally accommodates the circular rotation pattern of the center pivot irrigation system. Secondly, the microwave radiometers provide a spatial characterization of soil water content in the field, essentially offering a 2D representation of the observed data. In light of this, the 3D version of the Richards equation is well-suited to handle such spatial observations. The Richards equation in cylindrical coordinate form is expressed as~\cite{agyeman2021soil}:
\begin{multline}\label{eq:Richards_polar}
		\frac{\partial \theta_v}{\partial t} = C(h)\frac{\partial h}{\partial t}=\frac{1}{r}\frac{\partial}{\partial r}\bigg[r K(h)\frac{\partial h}{\partial r}\bigg]+\frac{1}{r}\frac{\partial}{\partial \theta}\bigg[\frac{K(h)}{r}\frac{\partial h}{\partial \theta}\bigg]+ \\ \frac{\partial}{\partial z}\bigg[K(h)\bigg(\frac{\partial h  }{\partial z}+ 1\bigg)\bigg]-S\left(h,z\right)
\end{multline}
where $r,\theta,z$ represent the radial, azimuthal, and axial  spatial variables, respectively. The term $h ~(\text{m})$ is the pressure head, $\theta_v ~(\text{m}^3 \text{m}^{-3})$ is the volumetric soil moisture content,  $t~(\text{s})$ is the temporal variable,  $K(h)~(\text{ms}^{-1})$ is the unsaturated hydraulic water conductivity, $ C(h)~(\text{m}^{-1})$ is the capillary capacity and $S\left(h,z\right)~(\text{m}^3 \text{m}^{-3}\text{s}^{-1})$  is the sink term which represents the root water extraction rate. 

{The term $S\left(h,z\right)$ is incorporated into Eq. (\ref{eq:Richards_polar}) to account for the interaction between plants and soil. This consideration is important as crops act as significant sinks, drawing moisture from the soil. The inclusion of $S\left(h,z\right)$ in Eq. (\ref{eq:Richards_polar}) is thus essential for accurately modeling the influence of plant root water uptake on soil moisture dynamics. In this study, the sink term proposed in~\cite{agyeman2021soil} is adopted, where the Feddes model is employed to represent $S\left(h,z\right)$. 
	Additionally, optimal root water uptake conditions are assumed in the field under investigation, leading to the use of a water stress reduction factor of 1 in the calculation of $S\left(h,z\right)$. This assumption is appropriate given the well-irrigated nature of the field and serves to minimize the influence of uncertain Feddes model parameters on both soil moisture state estimation and hydraulic parameter retrieval. Interested readers may refer to the aforementioned work for a comprehensive description of $S\left(h,z\right)$.}
The soil hydraulic functions $\theta_v (h)$, $K(h)$ and $C(h)$ in Eq. (\ref{eq:Richards_polar}) are described by the Mualem-van Genucthen model~\cite{van1980closed}:
\begin{figure}[!ht]
	\centering
	\centerline{\includegraphics[width=0.40\textwidth]{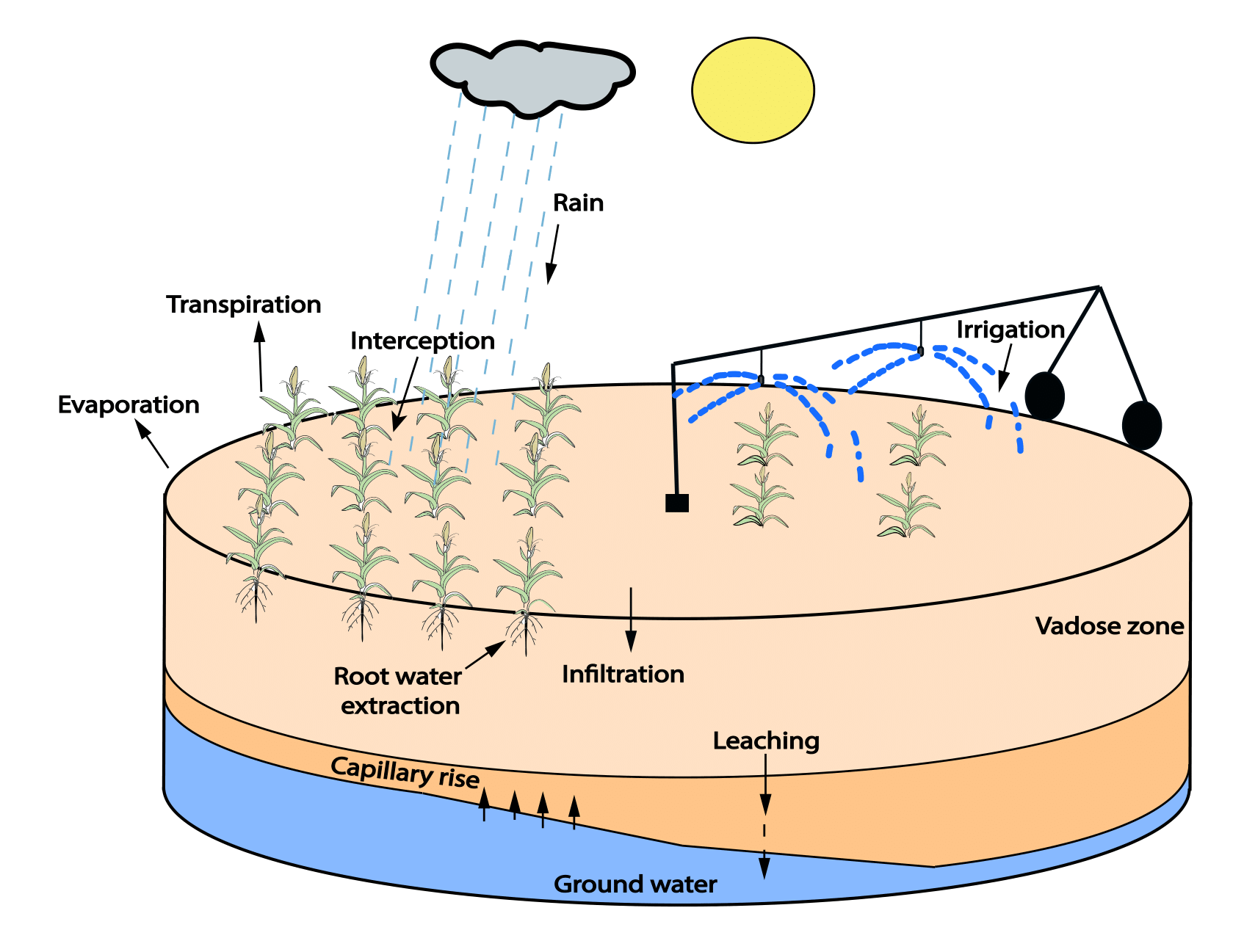}}
	\caption{An agro-hydrological system~\cite{agyeman2021soil}.}\label{fig:Polar_Agrohydrological}
\end{figure}
\begin{eqnarray}\label{eq:thetareln}
	\theta_v (h)=\theta_r +(\theta_s-\theta_r)\bigg[\frac{1}{1+(-\alpha h)^n}\bigg]^{1-\frac{1}{n}}
\end{eqnarray}
\begin{multline}
		K(h)=K_s\bigg[(1+(-\alpha h)^n)^{-\big(\frac{n-1}{n}\big)}\bigg]^{\frac{1}{2}}\times \\
	\Bigg[1-\bigg[1-\Big[(1+(-\alpha h)^n)^{-\big(\frac{n-1}{n}\big)}\Big]^{\frac{n}{n-1}}\bigg]^{\frac{n-1}{n}}\Bigg]^2
\end{multline}
\small 
\begin{equation}
	C(h)=\begin{cases}
		(\theta_s-\theta_r)~\alpha n~\bigg[1-\frac{1}{n}\bigg]~(-\alpha h)^{n-1}\big[1+(-\alpha h)^n\big]^{-\big(2-\frac{1}{n}\big)} & h <0\\
		{S_r} & h\geq 0 
	\end{cases}
\end{equation}
\normalsize
where $\theta_s~(\text{m}^3\text{m}^{-3})$, $\theta_r~(\text{m}^3\text{m}^{-3})$, $K_s~(\text{ms}^{-1})$, $S_r~ (\text{m}^{{-1}})$ are the saturated volumetric moisture content, residual moisture content, saturated hydraulic conductivity and specific storage coefficient, respectively and $n$, $\alpha$ are curve-fitting  soil hydraulic properties. The interested reader may refer to~\cite{agyeman2021soil} for a thorough description of the agro-hydrological system equipped with a center pivot irrigation system and the derivation of Eq. (\ref{eq:Richards_polar}).
\begin{remark}
	While a 1D version of the Richards equation could  in principle be employed in this work, especially in scenarios where the field is relatively flat, its application faces notable challenges within the context of this study’s observations and sensor configuration. Primarily, the 2D nature of the observations necessitates the deployment of multiple 1D models at each time step during the estimation process. This introduces practical complexities, including storage concerns due to the need to store error covariance matrices for each 1D model. Moreover, accommodating the rotational nature of the observations requires the estimation technique to dynamically switch between different 1D models, further complicating the computational process. Additionally, accurately capturing both the irrigated areas and the specific locations where measurements are taken at each time step presents significant challenges within a framework that employs a 1D version of the Richards equation. 
\end{remark}
\subsection{Numerical Solution of the Field Model} \label{sec:Numerical Model Development}
Eq. (\ref{eq:Richards_polar}) is solved numerically for the following boundary conditions:
\begin{alignat}{2}
	\frac{\partial h\left(r,\theta, z\right)}{\partial r}&=0 &&\text{at}~\left(r=0,~\theta,~z\right) \\
	\frac{\partial h\left(r,\theta, z\right)}{\partial r}&=0 &&\text{at}~\left(r=H_r,~\theta,~z\right) \label{eq:leftBC} \\ 
	\frac{\partial (h\left(r,\theta,z\right)+z)}{\partial z}&=1  &&\text{at}~\left(r,~\theta,~z=0\right) \\ 
	\frac{\partial (h\left(r,\theta,z\right))}{\partial z}&=-1-\frac{u_{\text{irr}}-\text{EV}}{K(h)} ~~~~&&\text{at}~\left(r,~\theta,~z=H_z\right)\label{eq:topBC} 
\end{alignat}
where $H_r$ in Eq. (\ref{eq:leftBC}) is the total radius of the field, $H_z$ in Eq. (\ref{eq:topBC}) is the length of the soil column. $u_{\text{irr}}$ (m/s) and $\text{EV}$ (m/s) in Eq. (\ref{eq:topBC}) represents the irrigation and the potential evapotranspiration rates, respectively.

The two-point central difference scheme proposed in \cite{agyeman2021soil} is adopted as the solution method of Eq. (\ref{eq:Richards_polar}) in this work. 
The interested reader may refer to~\cite{agyeman2021soil} for the full details of the mechanisms that can be used to guarantee a reliable numerical solution of Eq.~(\ref{eq:Richards_polar}). Note that the description of the potential evapotranspiration rates presented in~\cite{agyeman2021soil} are also adopted in this work. 
\subsection{State-space Representation of the Field Model}
After carrying out the numerical discretization, the field model can be expressed in state-space form as follows:
\begin{align}
	\dot{x} (t)&=\mathcal{F}(x(t),u(t), {\bm\Phi}) + \omega(t) \label{eq:statespace} \\ 
	y(t)&={{M}}(t)\mathcal{H}(x(t), \bm\Phi) + v(t) \label{eq:output}
\end{align}
where $x(t)\in \mathbb{R}^{N_x}$ represents the state vector containing $N_x$ pressure head values. The term $N_x$ is the total number of discretization nodes of the system and is the product of the total number of discretization points in the radial, azimuthal, and axial directions, denoted by $N_r$, $N_{\theta}$, and $N_z$, respectively.
$u(t)\in \mathbb{R}^{N_u}$ and $\omega (k)\in \mathbb{R}^{N_x}$ represent the inputs and the model disturbances, respectively. To model field soil texture heterogeneity, it is considered that each node has its own hydraulic parameters including $\{K_s, \theta_s, \theta_r, \alpha, n\}$. In Eqs.~(\ref{eq:statespace})-(\ref{eq:output}), $\bm\Phi\in\mathbb{R}^{N_p}$ represents the collection of all the soil hydraulic parameters of all the $N_x$ spatial node. 

In Eq.~(\ref{eq:output}), $y_k\in \mathbb{R}^{N_y}$, $v_k\in \mathbb{R}^{N_y}$ respectively denote the measurement vector and the measurement noise. Eq.~(\ref{eq:output}) is the general form of Eq.~(\ref{eq:thetareln})  since the microwave radiometers provide volumetric soil moisture observations. {The matrix $\mathcal{M}(t)$ serves as a selection matrix utilized to choose the states or nodes along the axial direction of the soil column that collectively contribute to the microwave observations at time step $t$. The depth of the soil column considered when constructing $\mathcal{M}(t)$ corresponds to the penetration depth of the microwave radiometers. 
	
	Note that the microwave radiometers measure the soil moisture content as the center pivot irrigates the field. Thus, the locations of the field at which the measurements are obtained change during the rotation cycle of the center pivot. This explains the explicit dependence of the selection matrix $\mathcal{M}(\cdot)$ on time $t$. $\mathcal{M}(t)$ is defined as {$\mathcal{M}(t)=\frac{1}{N_c}\begin{bmatrix}
			I_{N_y \times N_y}(t) & \emptyset_{N_y \times (N_x-N_y)}(t)
		\end{bmatrix}$}, where $N_c$ is the number of axial nodes in the penetration depth of the microwave sensors and $N_y$  is the number of soil water content measurements obtained at time $t$. The matrix $\mathcal{M}(t)$ contains both the identity matrix $I$ and the zero matrix $\emptyset$. However, it is worth noting that the positions of these matrices in $\mathcal{M}(t)$ are not constant. The exact placement of $I$ and $\emptyset$ in $\mathcal{M}(t)$ may change depending on the specific locations of the measurements at time instant $t$. }

\begin{remark}
	In this study, an equal weight of $\frac{1}{N_c}$ is allocated to the states within the penetration depth of the microwave radiometers, which collectively contribute to the soil moisture observations. It is plausible that states at different points within the penetration depth may have varying degrees of influence on the soil moisture observation. However, determining these exact weights has been noted to be challenging. Therefore, employing equal weights provides a practical solution in the absence of precise weight information. Nonetheless, if precise weights are available, they can be readily integrated into $\mathcal{M}(\cdot)$. It is anticipated that accurate weight determination would further improve the estimation accuracy of the proposed approach.
\end{remark}

\section{Proposed Simultaneous State and Parameter Estimation}
\label{ch3:sec:Proposed-Procedure}
As noted earlier, complete knowledge of the soil moisture distribution across the field is required for the design of a closed-loop irrigation system. To achieve this, the estimation framework is formulated to recover the full set of soil moisture states (all state variables of the field model). Regarding model parameters, a prior identifiability analysis is conducted to determine whether the chosen set of hydraulic parameters can be uniquely estimated from the available field observations. When full parameter identifiability is not attainable, a parameter selection step is introduced to isolate a subset of parameters that are most reliably estimable from the data. These selected parameters, referred to as the most estimable parameters, are then estimated jointly with the full soil moisture distribution.
Specifically, parameter identifiability is assessed using a sensitivity analysis approach, and the selection of estimable parameters is performed using the orthogonal projection technique. The joint estimation of soil moisture states and selected parameters is carried out using a modified version of the EKF algorithm.

Based on this overall framework, the remainder of this section is organized as follows. First, the sensitivity analysis method is described, followed by a procedure for constructing the sensitivity matrix in systems with spatially varying measurements. Next, the orthogonal projection technique for parameter selection is presented. The section concludes with a description of the modified EKF algorithm, adapted to accommodate changes in the set of estimable parameters over time.

\subsection{Sensitivity Analysis}
Sensitivity analysis evaluates how changes in a model's parameters affect its outputs, making it an essential tool for investigating parameter estimability. This involves generating trajectories that describe the relationship between variations in outputs and the model’s parameters, which are then used to form a sensitivity matrix. This matrix serves as the basis for analyzing the estimability of the model’s parameters. For the field model, the sensitivity of the system state \(x\) with respect to the parameters \(\bm{\Phi}\) is given by:
\begin{align}
	\frac{dx_{\bm{\Phi}}(t)}{dt} &= \frac{\partial \mathcal{F}}{\partial x}x_{\bm{\Phi}}(t)  +  \frac{\partial \mathcal{F}}{{\partial \bm{\Phi}}} \label{eq:sens_state}\\ 
	y_{{\bm{\Phi}}}(t) &= \frac{\partial \mathcal{H}}{\partial x}x_{\bm{\Phi}}(t)  +  \frac{\partial \mathcal{H}}{{\partial \bm{\Phi}}} \label{eq:sens_out}
\end{align}
where $x_{\bm{\Phi}}(t) = \frac{\partial x(t)}{ \partial \bm{\Phi}}$ and $y_{\bm{\Phi}}(t) = \frac{\partial y(t)}{ \partial \bm{\Phi}}$.

Equations~(\ref{eq:sens_state}) and~(\ref{eq:sens_out}) define the sensitivity equations, which are solved concurrently with the field model. The solution process requires an appropriate initial state for the field model, the nominal values of the parameter vector $\bm{\Phi}$, and an initial system state sensitivity set to \(x_{\bm{\Phi}}(t_0) = \bm{0}_{N_x \times N_p}\) (the zero matrix). This procedure is performed over a series of time steps, \(t_0, t_1, \dots, t_N\), with the condition that the total number of time steps must be at least equal to the number of parameters in the field model. At each time step \(t_k\), the instantaneous sensitivity matrix, \(s_y(t_k, \bm{\Phi})\), is constructed as follows:
\begin{equation}
	s_{y}(t_k,\bm{\Phi}) = \left[
	\begin{array}{cccc} 
		\frac{\partial y_{1}(t_k)}{\partial \Phi_{1}} & \frac{\partial y_{1}(t_k)}{\partial \Phi_{2}} & \cdots & \frac{\partial y_{1}(t_k)}{\partial \Phi_{N_p}}\\ 
		\frac{\partial y_{2}(t_k)}{\partial \Phi_{1}} & \frac{\partial y_{2}(t_k)}{\partial  \Phi_{2}} & \cdots & \frac{\partial y_{2}(t_k)}{\partial \Phi_{N_p}}\\ 
		\vdots &\vdots&\vdots&\vdots\\
		\frac{\partial y_{N_{y}}(t_k)}{\partial \Phi_{1}} & \frac{\partial y_{N_{y}}(t_k)}{\partial \Phi_{2}} & \cdots & \frac{\partial y_{N_{y}}(t_k)}{\partial \Phi_{N_p}}\\
	\end{array} 
	\right]
\end{equation}
The elements of \(s_y(t_k, \bm{\Phi})\) typically differ in physical dimensions and magnitudes, making it necessary to scale each element to enable a meaningful comparison of the impact of each parameter on the system’s outputs. This scaling produces a normalized instantaneous sensitivity matrix, \(\tilde{s}_y(t_k, \bm{\Phi})\), which is defined as:
\begin{equation}
	\tilde{s}_{y}(t_k,\bm{\Phi}) = \left[
	\begin{array}{cccc} 
		\frac{\Phi_1}{y_1(t_k)}\frac{\partial y_{1}(t_k)}{\partial \Phi_{1}} & \frac{\Phi_2}{y_1(t_k)}\frac{\partial y_{1}(t_k)}{\partial \Phi_{2}} & \cdots & \frac{\Phi_{N_p}}{y_1(t_k)}\frac{\partial y_{1}(t_k)}{\partial \Phi_{N_p}}\\ 
		\frac{\Phi_1}{y_2(t_k)}\frac{\partial y_{2}(t_k)}{\partial \Phi_{1}} & \frac{\Phi_2}{y_2(t_k)}\frac{\partial y_{2}(t_k)}{\partial  \Phi_{2}} & \cdots & \frac{\Phi_{N_p}}{y_2(t_k)}\frac{\partial y_{2}(t_k)}{\partial \Phi_{N_p}}\\ 
		\vdots &\vdots&\vdots&\vdots\\
		\frac{\Phi_1}{y_{N_y}(t_k)}\frac{\partial y_{N_{y}}(t_k)}{\partial \Phi_{1}} & \frac{\Phi_2}{y_{N_y}(t_k)}\frac{\partial y_{N_{y}}(t_k)}{\partial \Phi_{2}} & \cdots & \frac{\Phi_{N_p}}{y_{N_y}(t_k)}\frac{\partial y_{N_{y}}(t_k)}{\partial \Phi_{N_p}}\\
	\end{array} 
	\right]
\end{equation}
where \(y_{i}(t_k)\) represents the \(i^{\text{th}}\) observation at time step \(t_k\), and \(\Phi_{j}\) denotes the nominal value of the \(j^{\text{th}}\) parameter. The overall scaled sensitivity matrix, \(\tilde{S}_y(t_0, t_1, \dots, t_N, \bm{\Phi})\), is constructed by stacking the scaled instantaneous sensitivity matrices from each time step, resulting in:

\begin{equation}
	\tilde{S}_y(t_0,t_1,\dots,t_N,\bm{\Phi}) = \left[ 
	\begin{array}{c}
		\tilde{s}_y(t_0,\bm{\Phi})\\
		\tilde{s}_y(t_1,\bm{\Phi})\\
		\vdots\\
		\tilde{s}_y(t_{k},\bm{\Phi})\\
		\vdots\\
		\tilde{s}_y(t_N,\bm{\Phi})
	\end{array}
	\right]
\end{equation}

\subsubsection{Output Sensitivity Matrix for Spatially-Varying Measurements}
\label{ch3:sec:Sensitivity Analysis case 2}
\begin{figure}[t]
	\centering
	{\includegraphics[width=0.45\textwidth]{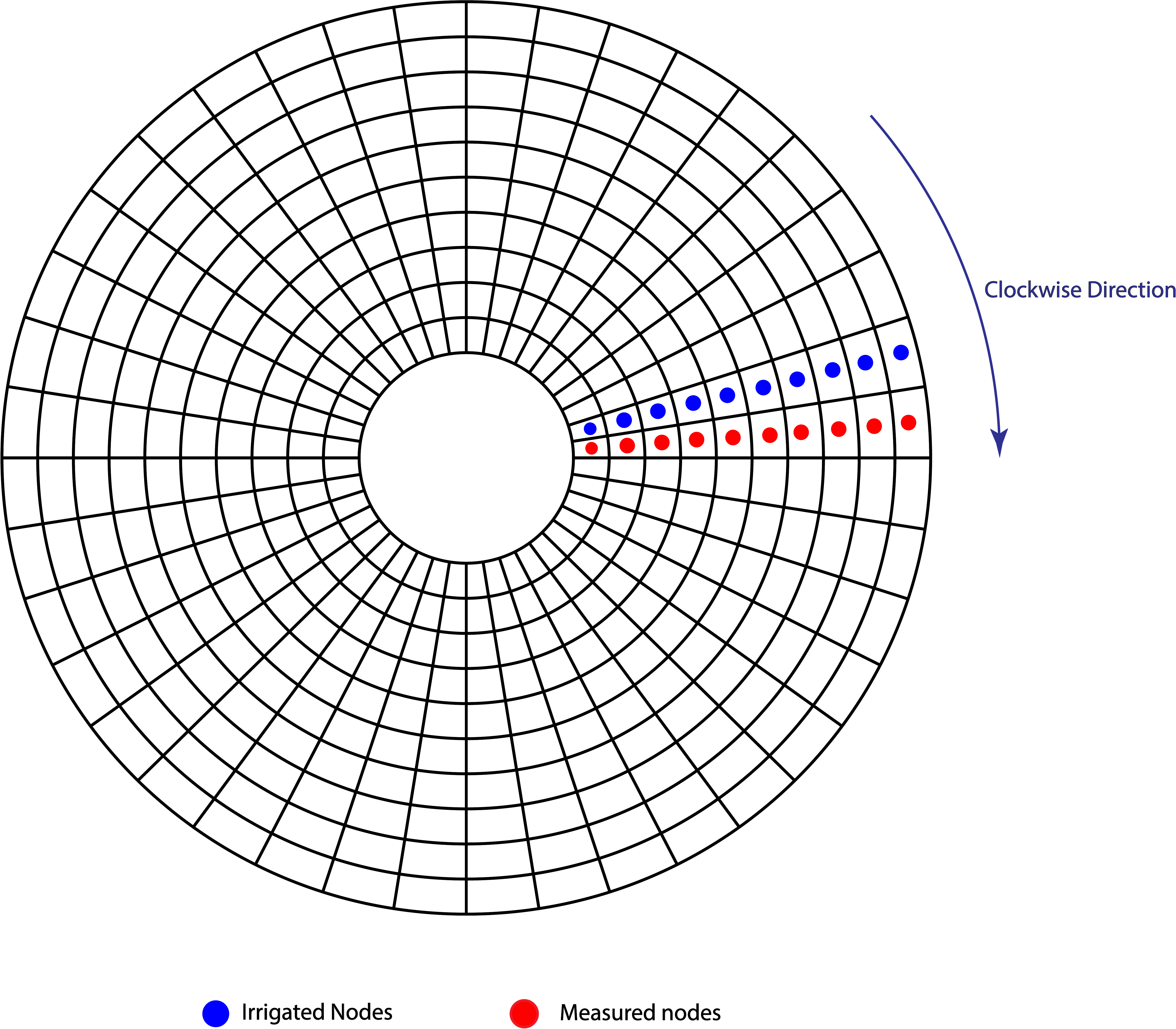}}
	\caption{{A schematic representation of center pivots equipped with microwave sensors.}}
	\label{fig:radiometer_pivot}
\end{figure}

As previously mentioned, the microwave remote sensors employed in this study measure soil moisture as the center pivot system rotates. The pivot’s movement causes the measurement locations to change continuously, resulting in soil moisture observations that vary across both space and time. Figure~\ref{fig:radiometer_pivot} illustrates the operation of a microwave remote sensor mounted on a center pivot system. Typically, these sensors capture the moisture content of areas that have not yet been irrigated. In Figure~\ref{fig:radiometer_pivot}, the blue dots represent the nodes (or field locations) being irrigated at time step \(t_k\), while the red dots correspond to the nodes measured by the sensor. As the pivot moves to a new location at time step \(t_{k+1}\), the measured nodes also shift accordingly.

The use of spatially varying measurements necessitates adjustments in the analysis and construction of the overall scaled sensitivity matrix. Aggregating all scaled instantaneous sensitivity matrices is not appropriate because these matrices are derived from measurements at different field locations across consecutive time steps. To account for spatial variability, the construction of the overall scaled sensitivity matrix should include only the instantaneous sensitivity matrices associated with measurements taken at the same spatial locations. As a result, multiple overall scaled sensitivity matrices are constructed, each corresponding to a specific field location where measurements are collected. 

In the context of the sensor configuration shown in Figure~\ref{fig:radiometer_pivot}, this requires constructing overall scaled sensitivity matrices for each node in the azimuthal direction of the field model. To clarify the methodology, consider an example where the field model is divided into 40 nodes along the azimuthal direction, and sensitivity analysis spans consecutive rotations of the center pivot. For the first azimuthal node, only the scaled instantaneous sensitivity matrices at time steps \(t_k = 1, 41, 81, 121, \dots\) are included in constructing its overall scaled sensitivity matrix. Specifically, the overall scaled output sensitivity matrix for the first azimuthal node, denoted as \(\tilde{S}^{1}_y(t_0, t_1, \dots, t_N, \bm{\Phi})\), is defined as:
\begin{equation}
	\tilde{S}^{1}_y(1,41,81,\dots,\bm{\Phi}) = \left[ 
	\begin{array}{c}
		\tilde{s}_y(1,\bm{\Phi})\\
		\tilde{s}_y(41,\bm{\Phi})\\
		\tilde{s}_y(81,\bm{\Phi})\\
		\vdots
	\end{array}
	\right]
\end{equation}
For the second azimuthal node, only the scaled instantaneous sensitivity matrices corresponding to the time steps \(t_k = 2, 42, 82, 122, \dots\) are utilized to construct its overall scaled output sensitivity matrix. This matrix, representing the second azimuthal node, is denoted as \(\tilde{S}^{2}_y(t_0, t_1, \dots, t_N, \bm{\Phi})\), and is defined as:
\begin{equation}
	\tilde{S}^{2}_y(2,42,82,\dots,\bm{\Phi}) = \left[ 
	\begin{array}{c}
		\tilde{s}_y(2,\bm{\Phi})\\
		\tilde{s}_y(42,\bm{\Phi})\\
		\tilde{s}_y(82,\bm{\Phi})\\
		\vdots
	\end{array}
	\right]
\end{equation}
This procedure is repeated for all remaining azimuthal nodes, resulting in the construction of an overall scaled sensitivity matrix for each node. To assess the estimability of parameters for each azimuthal node, the Singular Value Decomposition (SVD) technique is applied to the corresponding overall scaled sensitivity matrix. For any given azimuthal node, a full column rank of the overall scaled sensitivity matrix serves as a sufficient condition for parameter estimability~\cite{walter2013identifiability}.

\subsection{Orthogonal Projection for Spatially-Varying Measurements}
When the overall scaled sensitivity matrix is rank-deficient, it indicates that not all parameters in \(\bm{\Phi}\) can be uniquely estimated from the available observations. To obtain reliable parameter estimates, it becomes necessary to identify a subset of estimable parameters from \(\bm{\Phi}\). Orthogonal projection is used for parameter selection, as it enables dimensionality reduction while preserving the physical significance of the selected parameters--unlike approaches such as PCA.

The orthogonal projection technique aims to determine an orthogonal basis by ranking the columns of the scaled sensitivity matrix based on their norms and the linear dependence of the sensitivity vectors. For \(\tilde{S}^{i}_y(\cdot)\), the overall scaled sensitivity matrix for azimuthal node \(i\), the steps used to identify the estimable hydraulic parameters are summarized as follows~\citep{yao2003modeling}:
\begin{enumerate}
	\item   Evaluate the norm of each column of the scaled $\tilde{S}^{i}_y(\cdot)$, initialize $j = 1$ and select the column $X_j$ with the largest norm.
	\item   Estimate the information $Z_j$ in $\tilde{S}^{i}_y(\cdot)$, expressed by $X_j$, and calculate the residual matrix $R_j$.
	\begin{equation}
		Z_j \coloneqq X_j(X_j^{T}X_j)^{-1}X_j^{T}\tilde{S}_y^i(\cdot)
	\end{equation}
	\begin{equation}
		R_j \coloneqq \tilde{S}^{i}_y(\cdot) - Z_j
	\end{equation}
	\item  Evaluate the norm of each column of the residual matrix $R_j$; select the column from $S_y^i(\cdot)$ that corresponds to the column with the largest norm in $R_j$ and add the selected column to $X_j$ to form $X_{j+1}$.
	
	\item  If the rank$\left(X_{j}\right)$ = rank$\tilde{S}^{i}_y(\cdot)$ or the largest norm of the columns of $R_j$ is smaller than a prescribed cut-off value, terminate the algorithm and the selected elements of $\bm{\Phi}$ correspond to the selected columns in $X_j$; otherwise, repeat Step 2 to Step 4 with $j \gets j+1$.
\end{enumerate} 
The orthogonal projection method should be applied to all the overall scaled sensitivity matrices constructed for the azimuthal nodes of the field model. To simplify the online estimation of soil moisture and selected hydraulic parameters, the identification of estimable soil hydraulic parameters can be performed offline through extensive simulations. It is essential to recognize that the set of estimable parameters may vary depending on the azimuthal node. For example, when measurements are taken from azimuthal node \(i\), the corresponding estimable parameters may differ from those associated with measurements taken at another azimuthal node \(j\) (\(j \neq i\)). All parameters identified as potentially estimable across the rotation of the pivot are compiled into a single vector, denoted by \(\bm{\Phi}^e\), while the remaining parameters are grouped into \(\bm{\Phi}^{ne}\).

\subsection{Simultaneous State and Parameter Estimator}~\label{sec:State and Parameter Estimator}
Before designing the estimator, the estimable parameters are incorporated into the original states of the field model to create an augmented system, \(x_a(t) = [x(t)~~\bm{\Phi}^e]\). The continuous-time system is then converted into its discrete-time equivalent. The discrete-time representation of the augmented system, including the effect of the non-estimable parameters, is expressed as:
\begin{align} 
	\label{eq:discrete model}
	x_{a}(k+1) &= \mathcal{F}_{a}(x_{a}(k),u(k), \bm{\Phi}^{ne}) + \omega_{a}(k)\\
	y(k) &= {{M}_a}(k)\mathcal{H}_{a}(x_{a}(k), \bm{\Phi}^{ne}) + v(k)
\end{align} 
It is important to highlight that nominal values are assigned to the non-estimable parameters during the estimation of \(x_a\). However, at each sampling time, the elements of \(\bm{\Phi}^{ne}\) will be updated using the Kriging interpolation method, based on the estimated values of the elements in \(\bm{\Phi}^{e}\).

The EKF is employed for data assimilation in this framework. A detailed justification for its selection is provided in Remark~\ref{rm:ekf}.  The EKF is initialized with an initial estimate of the augmented state, \(\hat{x}_{a}(0)\), and its covariance matrix, \(P_{a}(0|0)\). During the prediction step, the augmented state and its covariance matrix are forecasted using the field model as follows:
\begin{equation}
	\label{eq:prediction}
	\hat{x}_{a}(k+1|k)=\mathcal{F}_{a}(\hat{x}_{a}(k|k),u(k),\bm{\Phi}^{ne})
\end{equation}
\begin{equation}\label{eq:propagation}
	P_{a}(k+1|k) = A(k) P_{a}(k|k) A(k)^T + Q
\end{equation}
where $A(k)=\frac{\partial \mathcal{F}_{a}}{\partial x_{a}}\big|_{\hat{x}_{a}(k|k),~u(k)}$ and $Q$ is the covariance matrix of the process disturbance $\omega_{a}$. 

In the update step of EKF, the predicted augmented state and its covariance matrix are updated using the observation $y(k+1)$ at time $k+1$ by:
\small
\begin{multline}
	\hat{x}_{a}(k+1|k+1)=\hat{x}_{a}(k+1|k) +  G(k+1)[y(k+1)-{\mathcal{M}_a}(k)\mathcal{H}_{a}(\hat{x}_{a}(k+1|k),\bm{\theta}^{ne})]
\end{multline}
\normalsize
\begin{equation}
	P_{a}(k+1|k+1)=[I-G(k+1)C(k+1)]P_{a}(k+1|k)
\end{equation}
In the above equations, $G(k+1)$ is the Kalman gain matrix that can be calculated as:
\small 
\begin{equation}
	\begin{aligned}
		G(k+1) = P_{a}(k+1|k) C^T(k+1)[C(k+1)P_{a}(k+1|k)C^T(k+1) + R]^{-1}
	\end{aligned}
\end{equation}
\normalsize
where $C(k+1)= M_a(k)\frac{\partial \mathcal{H}_{a}}{\partial x_{a}}\big|_{\hat{x}_{a}(k+1|k)}$ and $R$ is the covariance matrix of the measurement noise $v$.

As previously discussed, the set of estimable hydraulic parameters may vary over time. In the data assimilation process, the vector \(\bm{\Phi}^e\) encompasses all parameters identified as estimable through sensitivity analysis for each azimuthal node. At any given time step, if a parameter in \(\bm{\Phi}^e\) is not selected for estimation, the corresponding entries in the \(A(k)\) and \(C(k)\) matrices are set to zero. This adjustment accounts for the fact that the parameter is inactive for estimation during that specific time step. By employing this approach, the data assimilation process can dynamically adjust to changes in the set of estimable parameters while eliminating the need to handle multiple covariance matrices, thereby enhancing computational efficiency.

\begin{remark}\label{rm:ekf}
	The EKF was selected as the data assimilation technique due to its favorable trade-off between accuracy and computational efficiency. Although the Richards equation is highly nonlinear, the EKF provided reliable estimation performance in this application at a lower computational cost than the EnKF. In particular, a symbolic computing framework known as CasADi~\cite{andersson2019casadi} was used to compute the required Jacobian matrices, enabling efficient implementation of the EKF. Furthermore, in the real field case study, the initial state estimate was highly uncertain, necessitating the use of a large initial covariance matrix. While this is generally feasible within the EKF framework, it can pose challenges for the EnKF. Specifically, sampling ensembles from a distribution with high variance may yield implausible states that the model cannot propagate, potentially destabilizing the assimilation process~\cite{chirico2014kalman}. Nevertheless, in scenarios involving larger domains or cases where more accurate initial estimates of the system state are available, the EnKF may offer improved scalability and performance. In such settings, the EnKF could be substituted for the EKF within the proposed state and parameter estimation framework.
\end{remark} 

\section{Simulated Case Study} 
\label{ch3:sec:Results}
This section evaluates the performance of the proposed framework using simulated microwave sensor measurements. Initially, the system on which the simulations are based is described. Subsequently, a set of criteria is presented for evaluating the effectiveness of the proposed state and parameter estimator. Results of the sensitivity analysis and hydraulic parameter selection are also provided. Finally, the states estimation results  of some selected states in the field model are presented to emphasize the benefits of the proposed method.

\subsection{System Description and Simulation Settings}
\label{sec:Scenario 2}
This simulation experiment involved a field with a radius of 290 m and a depth of 0.3 m. The entire system was discretized into \(N_x = 12,000\) nodes, comprising 30 equally spaced nodes in the radial direction, 40 in the azimuthal direction, and 10 in the axial direction. Extensive simulations revealed that further mesh refinement in any direction did not significantly affect the state trajectories. Additionally, several simulation experiments revealed that a time step size of 6 minutes was suitable for temporal discretization. 
The study considered a heterogeneous field, with the spatial distribution of nominal hydraulic parameters shown in Figure~\ref{fig:soil_pars_sens_analyis} in Appendix~\eqref{sec:hyd_pars_sens_analysis}. Further more, a detailed description of the approach used to obtain the nominal parameters is also provided in Appendix~\eqref{sec:hyd_pars_sens_analysis}.

The initial pressure head for all nodes in the field model was randomly assigned values between \(-1.5~\text{m}\) and \(-1.35~\text{m}\). The center pivot operated at a speed of \(0.011~\text{m/s}\), delivering a constant irrigation rate of \(3.6~\text{mm/day}\) and irrigating the field during the first 8 hours of each day. A healthy barley crop in its development stage was assumed, with crop coefficient values ranging between 0.75 and 0.96. Daily reference evapotranspiration inputs of \(1.5~\text{mm/day}\), \(1.90~\text{mm/day}\), \(0.6~\text{mm/day}\), \(0.8~\text{mm/day}\), and \(2.40~\text{mm/day}\) were used in the simulations. Process noise and measurement noise were incorporated, with zero means and standard deviations of \(1 \times 10^{-6}~\text{m}\) and \(1 \times 10^{-2}~\text{m}^3\text{m}^{-3}\), respectively.

At each sampling time, 30 measurements were used to update elements of the augmented system in the EKF's correction step. The EKF was initialized as follows: \(\hat{x}_a(0) \in [1.10x_a(0), 1.15x_a(0)]\), where \(\hat{x}_a(0)\) was the initial estimate of the augmented system, and \(x_a(0)\) represented the actual initial state of the augmented system.

\subsection{Evaluation criteria}
The effectiveness of the proposed method was evaluated through three different cases. The first case (Case 1) involved only soil moisture estimation while accounting for uncertainty in the soil hydraulic parameters. In the second case (Case 2), all soil moisture states and hydraulic parameters were estimated without performing sensitivity analysis or selecting estimable parameters. The third case (Case 3) applied the proposed method to simultaneously estimate both soil moisture and hydraulic parameters, incorporating sensitivity analysis and parameter selection.

To quantify the performance of each case, the root mean square error (RMSE) at individual time steps and the average RMSE over the simulation period were used as evaluation metrics. The mathematical definitions of these metrics are provided as follows:
\begin{equation}
	RMSE_{x_{a}}(k) = \sqrt{\frac{\sum_{i=1}^{n_{x_{a}}}(\hat{x}_{a,i}(k) - x_{a,i}(k))^{2}}{n_{x_{a}}}}
\end{equation}
\begin{equation}
	RMSE_{x_{a}} = \frac{\sum_{k=0}^{N_{sim}-1}RMSE_{x_{a}}(k)}{N_{sim}}
\end{equation}
where RMSE$_{x_{a}}(k)$ with $k = 0, \cdots, N_{sim}-1$ shows the evolution of the RMSE value over time and RMSE$_{x_{a}}$ shows the average value. The term $N_{sim}$ is the total number of time steps in the simulation period.

\subsubsection{Determination of Significant Parameters}
\label{sec:Determination of Significant Parameters 2}
In this simulated case study, a sensing depth of 0.1 m was considered. Given the discretization of the field model, this depth corresponded to 1,200 nodes falling within the penetration range of the simulated microwave radiometers. Consequently, the model included a total of \(5 \times 1,200 = 6,000\) hydraulic parameters. Ideally, all model parameters would be included in the sensitivity analysis. However, preliminary simulation results indicated that parameters associated with nodes located beyond the sensing depth were not practically estimable from the available measurements. This prior insight was incorporated into the identifiability analysis by restricting the construction of the sensitivity matrix to only those parameters corresponding to nodes within the sensing depth. 
The parameters of the field model, \(\bm{\Phi}\), were represented as:
\[
\bm{\Phi} = [K_{s_1}, \theta_{s_1}, \theta_{r_1}, \alpha_{1}, n_{1}, \cdots, K_{s_{1200}}, \theta_{s_{1200}}, \theta_{r_{1200}}, \alpha_{1200}, n_{1200}]^{T}
\]
For each azimuthal node, the sensitivity matrix consisted of 6,000 columns, corresponding to the 6,000 model parameters. By performing singular value SVD on each sensitivity matrix, 6,000 singular values, denoted as \(\sigma_{1}, \cdots, \sigma_{6000}\), were obtained.
\begin{figure}[t]
	\centering
	\subfigure[Sector  1.]{
		\includegraphics[width=0.5\textwidth]{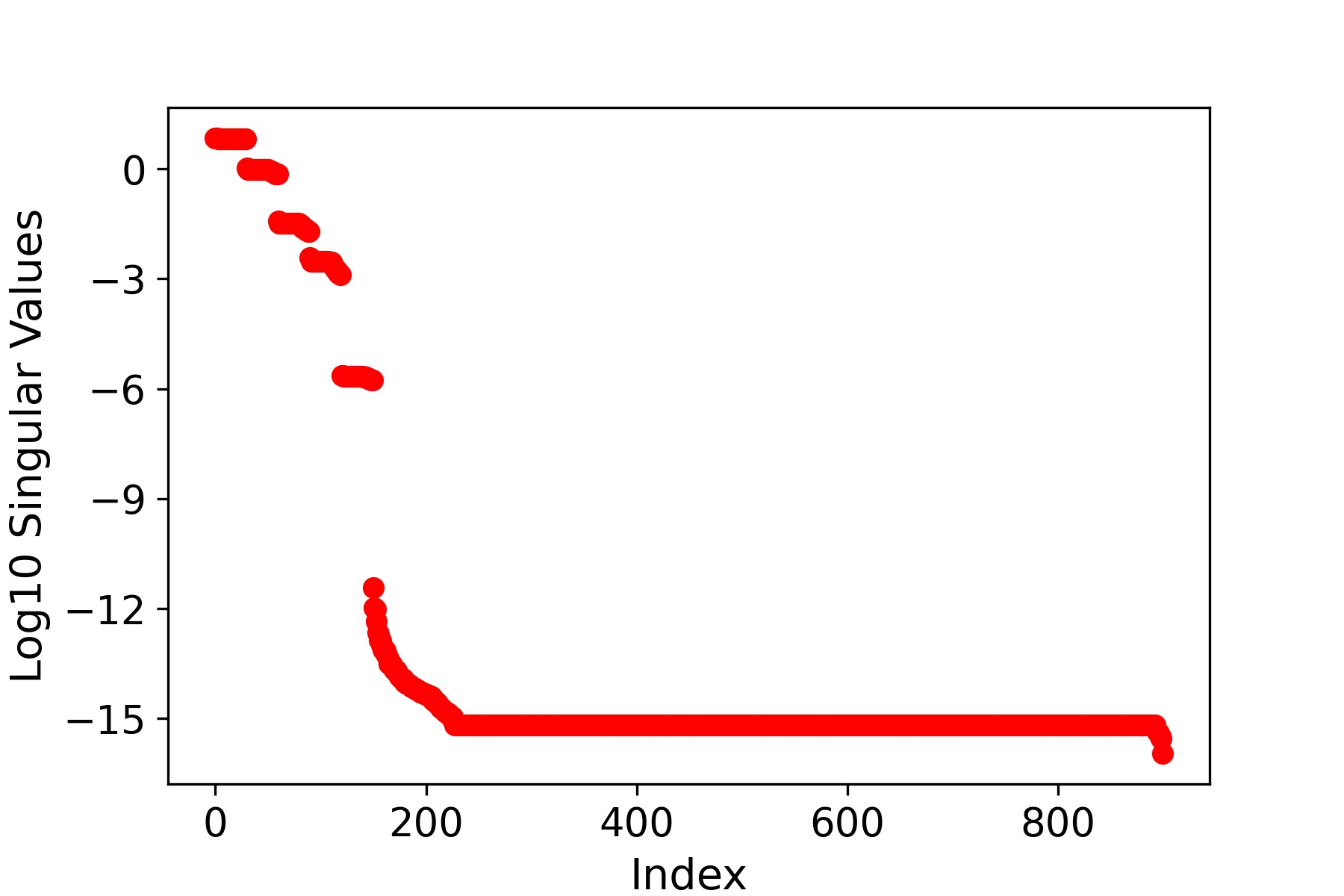}}%
		\vspace{-2mm}
	\subfigure[Sector 25.]{
		\includegraphics[width=0.5\textwidth]{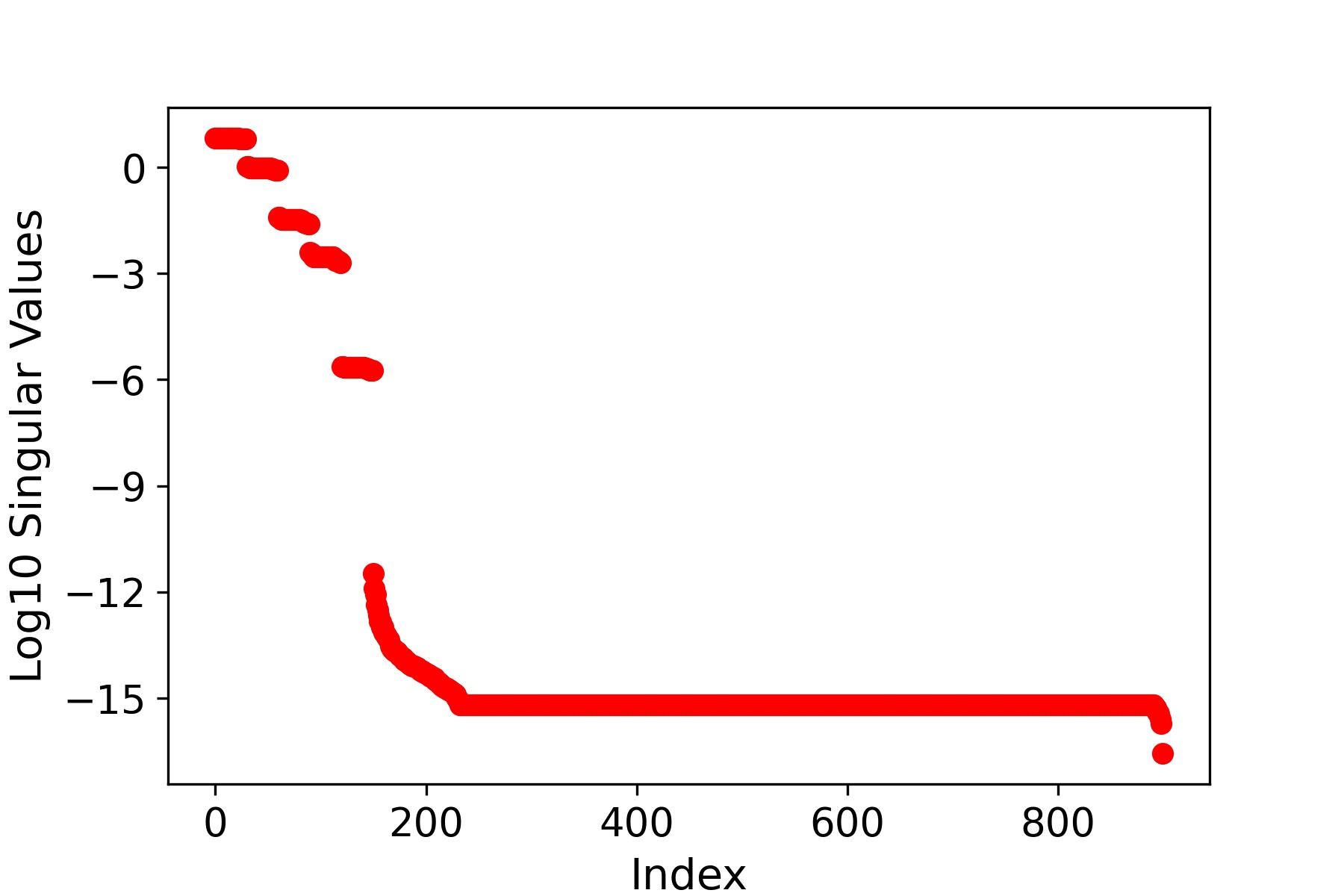}}
	\caption{Plots of the logarithm of the singular values of scaled output sensitivity matrix for some selected sectors.}
	\label{fig:sensitivity2}
\end{figure}
%
Figure~\ref{fig:sensitivity2} illustrates the largest 900 singular values of the overall sensitivity matrices for two sectors of the field, presented on a logarithmic scale. Each subplot in Figure~\ref{fig:sensitivity2} reveals a notable gap between \(\sigma_{150}\) and \(\sigma_{151}\), spanning approximately 5.8 decades on the logarithmic scale. This gap suggests a true zero value for the \(150^{\text{th}}\) singular value, indicating a potential lack of estimability. Consequently, the sensitivity matrix in the three depicted sectors is determined to have a rank of 150. This implies that, theoretically, only 150 of the 6,000 parameters can be uniquely estimated using the measurements taken at the specified sampling times. Similar results were observed for the remaining sectors of the field.

Following the parameter estimability analysis, a parameter selection step was conducted, using the rank of the sensitivity matrix as the termination criterion for the orthogonal projection method. Through this process, the orthogonal projection method identified the five hydraulic parameters---\(K_{s}\), \(\theta_{s}\), \(\theta_{r}\), \(\alpha\), and \(n\)---associated with the measured nodes of the field at each sampling time \(k\) as the most estimable parameters. Consequently, during each sampling period in the soil moisture and parameter estimation step, the proposed approach estimated the complete soil moisture profile of the field model along with the five hydraulic parameters corresponding to the 30 measured nodes.

\subsubsection{Results and Discussion}
\label{sec:Performance of EKF 2} 
\begin{figure}
	\centering
	\subfigure[Node 10700 at depth = 3.3 cm. ]{
		\includegraphics[width=0.42\textwidth]{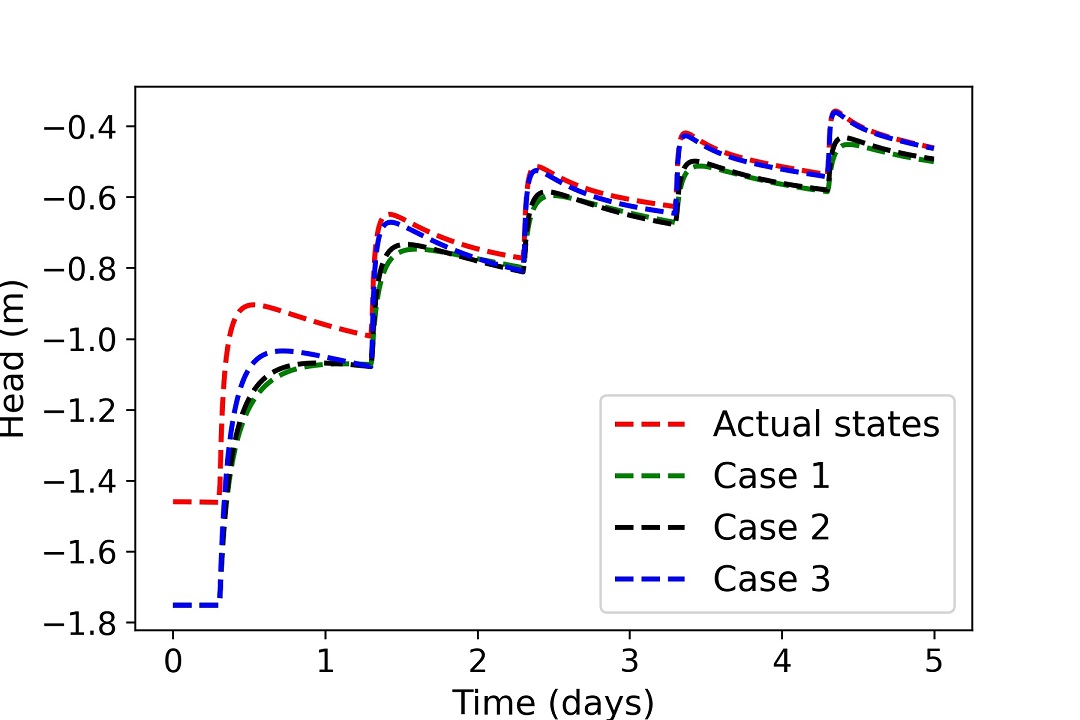}}
	\subfigure[Node 6601 at depth = 13.3 cm]{
		\includegraphics[width=0.42\textwidth]{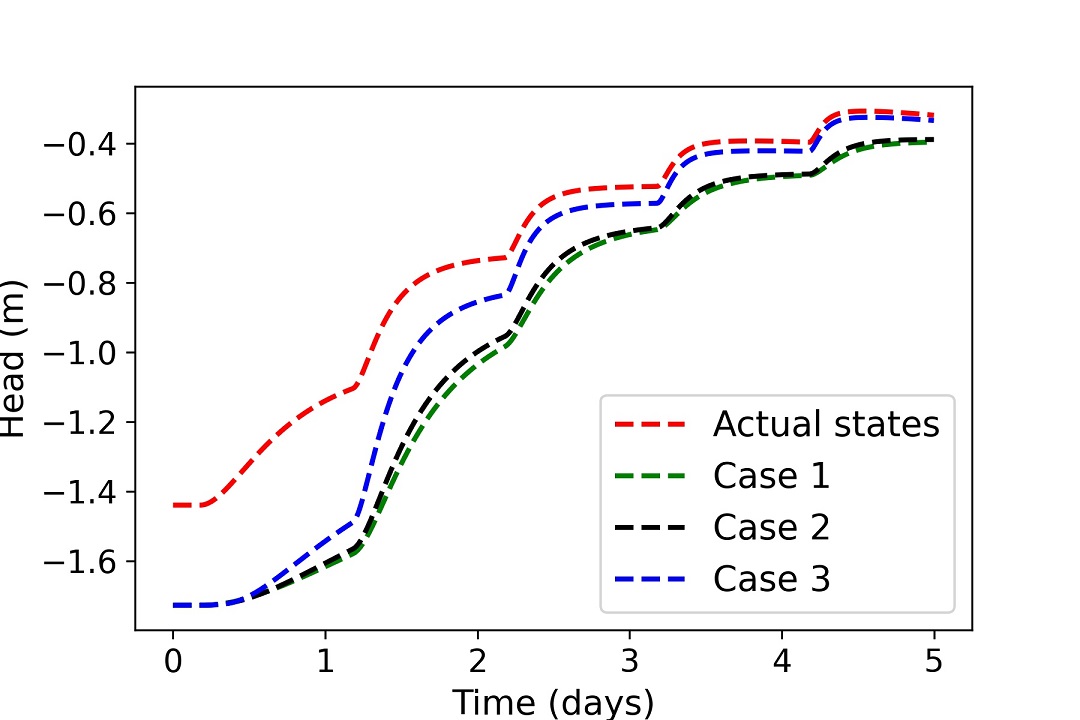}}
	\centering
	\subfigure[Node 4150 at depth = 20 cm.]{
		\includegraphics[width=0.42\textwidth]{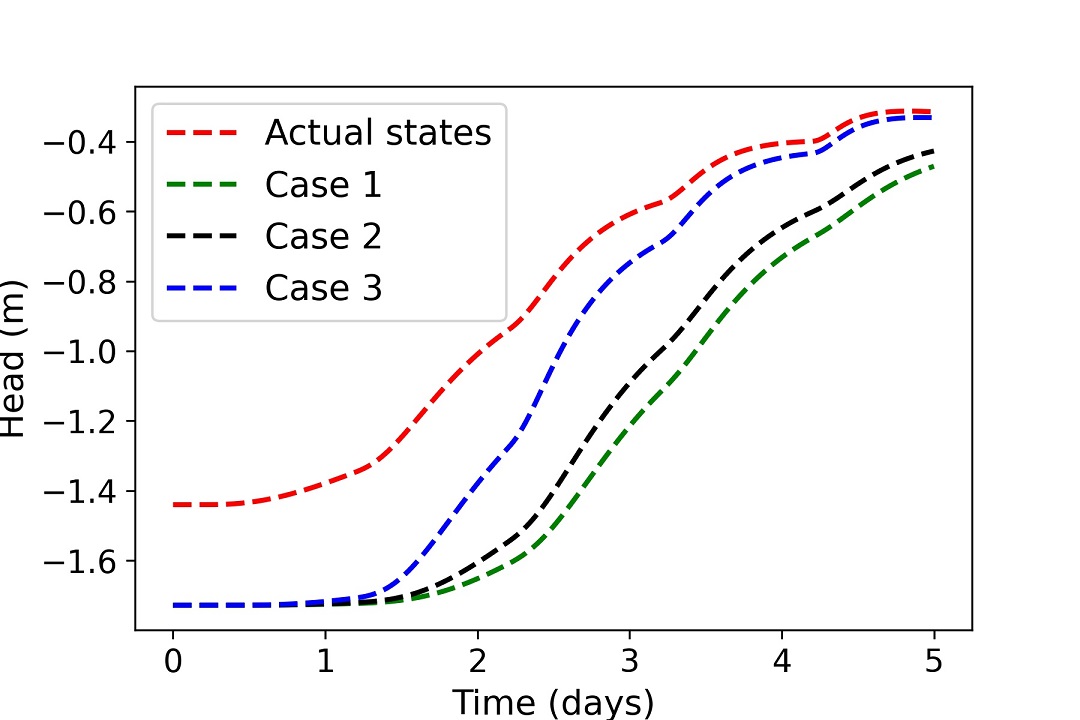}}
	\subfigure[RMSE]{
		\includegraphics[width=0.42\textwidth]{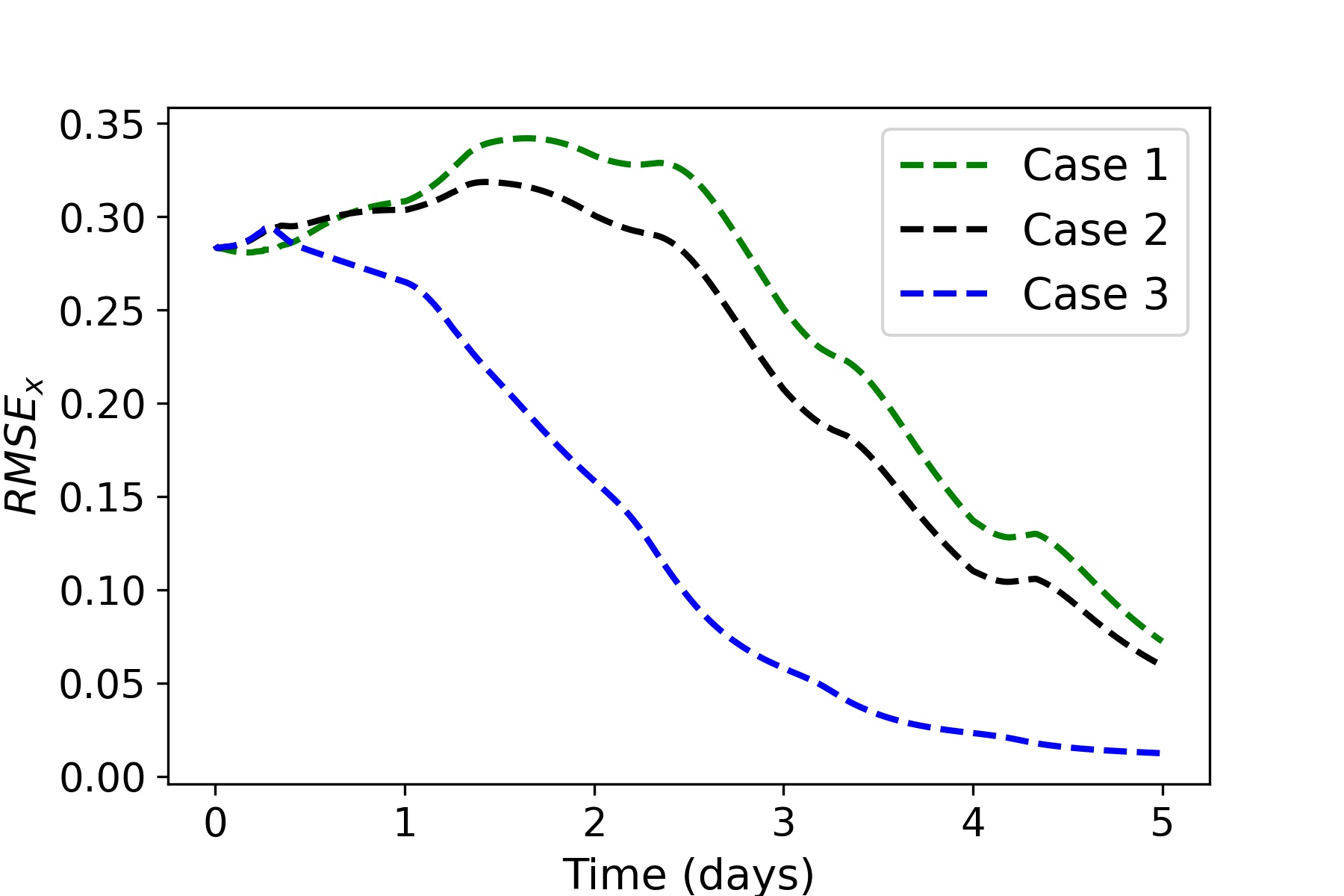}}
	\caption{(a)-(c) Some selected trajectories of the actual states (red lines), estimated states in Case 1 (green lines), estimated states in Case 2 (black lines), and estimated states in Case 3 (blue lines). (d) Evolution of the RMSE of the original state vector during the simulation time in Case 1 (red lines), Case 2 (blue lines) and Case 3 (black lines).}
	\label{fig:result2}. 
\end{figure}

Figure~\ref{fig:result2} illustrates the estimation results for selected nodes in the field model across the simulation period for the three cases. As anticipated, states closer to the soil surface exhibit greater variability compared to those located deeper within the soil profile. Among the three cases, Case 3 demonstrated the highest agreement with the actual states, followed by Case 2 and Case 1. The RMSE values computed for these cases, summarized in Table~\ref{tbl:rmse2}, confirm that the proposed approach provides the most accurate state estimates. 
The enhanced accuracy in soil moisture estimates in Cases 2 and 3 compared to Case 1 is consistent with prior findings~\cite{bo2020parameter,bo2020decentralized, medina2014kalman}, which highlight improved accuracy when soil hydraulic parameters are estimated alongside soil moisture. Furthermore, the superior estimation performance of Case 3 over Case 2 aligns with studies on state and parameter estimation~\cite{liu2021simultaneous,chis2011structural}, demonstrating that estimating non-estimable parameters alongside estimable parameters can degrade state estimation accuracy.
While the Case 3 results are comparable to other algorithms for soil moisture and parameter estimation~\cite{medina2014kalman,medina2014kalmanp3}, the anticipated improvement in accuracy stems from the preliminary parameter identifiability analysis conducted prior to the estimation study. This critical step, as emphasized in~\cite{liu2007uncertainty}, ensures effective and computationally efficient estimation of hydraulic parameters. Additionally, the proposed approach has the potential to reduce the computational complexity associated with joint estimation by systematically selecting a subset of hydraulic parameters for estimation during the parameter identifiability step.
However, the benefits of the proposed approach depend on conducting comprehensive parameter identifiability studies over a wide range of parameter realizations, which can be time-intensive. These studies must also account for diverse initial soil water conditions and other relevant factors, such as irrigation rates and weather conditions, as these significantly influence the results of the parameter identifiability analysis.
\begin{table}[t]
	\caption{A summary of the evaluation metrics for the 3 cases.}
	\renewcommand{\arraystretch}{1.5}
	\footnotesize
	\centering
	\begin{tabular}{cccc}
		\hline
		\hline
		Case & $\text{RMSE}_{x}~(\%)$ & $\text{RMSE}_{\bm{\theta}}~(\%)$ & $\text{RMSE}_{x_{a}}~(\%)$ \\
		\hline
		1 & 26.29 & n/a  &  26.29 \\
		\hline
		2 & 24.19 & 14.20 & 24.00\\
		\hline
		3 & 16.60 & 13.90 & 16.44\\
		\hline
	\end{tabular} \label{tbl:rmse2}
\end{table}

\section{Real Case Study}
In this section, the utility and performance of the proposed method are demonstrated using microwave remote sensor measurements obtained from a field equipped with a center pivot irrigation system. The section begins with a description of the study area, followed by an description of the microwave radiometers that are employed in this work. Following this, the numerical modeling of the investigated field is presented, accompanied by a series of data pre-processing steps designed to obtain an appropriate data representation for the estimator. Subsequently, the section explores the sensitivity analysis and the application of orthogonal projection approaches to the field model, reporting the primary outcomes of these steps. The design of the estimator is then introduced, along with a description of the criteria employed to evaluate the performance of the proposed approach. Finally, the section concludes by presenting and discussing the results derived from the investigation.

\subsection{Study Area}
\begin{figure}[t]
	\centerline{\includegraphics[width=0.35\textwidth]{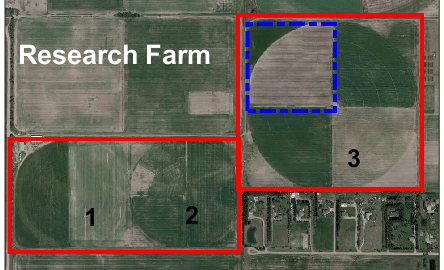}}
	\caption{Layout of the Research Farm operated by Lethbridge Polytechnic.} \vspace{2mm}
	\label{fig:layout_of_the_alberta_irrigation_center_}
\end{figure}

The study was conducted in 2021 at a Research Farm operated by Lethbridge Polytechnic, which is located at 49.7230 \textdegree N and 112.8001 \textdegree W, east of the City of Lethbridge in Alberta, Canada. The center has an approximate area of 0.81 km² and an average elevation of 888 m. The primary soil type in the area is clayey loam with a few sand lenses.  The layout of the center, delineated with the red rectangular blocks, is shown in Figure~\ref{fig:layout_of_the_alberta_irrigation_center_}. All three circular fields at the center are equipped with a center pivot irrigation system.
For the numerical investigation, a quadrant of Field 3 was selected, which is delineated by a blue dash-dotted rectangular block in Figure~\ref{fig:layout_of_the_alberta_irrigation_center_}. The soil moisture observations used in the study were obtained from three dual-polarization passive microwave radiometers/sensors that were mounted at an angle of 30\textdegree~on the center pivot of Field 3. 

\subsection{Microwave Remote Sensing Approach}
\begin{figure}[!ht]
	\centerline{\includegraphics[width=0.35\textwidth]{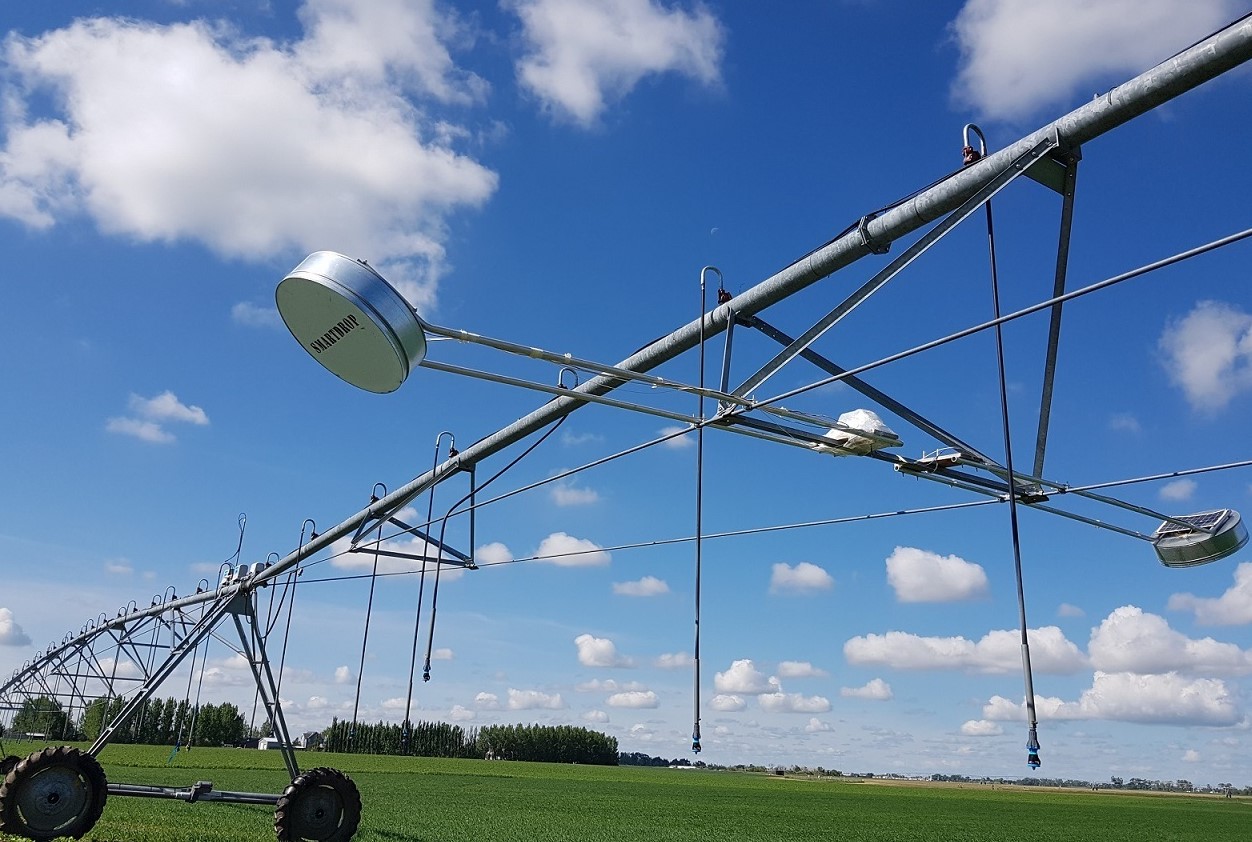}}
	\caption{Microwave radiometers mounted on a center pivot irrigation system.} \vspace{2mm}
	\label{fig:center_pivot_microwave_rad}
\end{figure}
The sensors were positioned at a height of 3 meters, each equipped with an antenna featuring a half-power beamwidth of 65\textdegree. This particular installation height, combined with the antenna's beamwidth, resulted in a nearly circular antenna footprint, spanning approximately 4 meters in diameter. The antenna dimensions measure 45 $\times$ 45 $\times$ 15 cm. These radiometers operate within a frequency range between 380 and 420 MHz, and they posses an effective sensing depth of 20 cm. Figure~\ref{fig:center_pivot_microwave_rad} depicts a center pivot irrigation system equipped with microwave radiometers.

Upon detection of movement within the center pivot irrigation system by its internal Global Positioning System (GPS), each radiometer is activated. The sensor's dual-polarized patch panel is oriented with respect to the soil to capture horizontal (H) and vertical (V) polarizations of the signal. The received H and V polarizations undergo a Fast Fourier Transform  process, converting the signals from the time domain into the complex spectra domain. Subsequently, the auto-correlation  and cross-correlation products are computed based on the horizontal and vertical polarization components derived from the complex spectra.

These correlation products are then converted into Stokes parameters (I, Q, U, and V). Parameter I represents the total power signal, while Q and U are the linearly polarized signals, and V contains any circularly polarized signals. The radiometer derives the radiometric power of the unpolarized component, which is the signal of interest, by subtracting the polarized components (Q, U, and V) from the total power. This radiometric power is further converted into brightness temperature $\text{T}_{\text{sys}}$ in Kelvin, utilizing the principles of the Dicke radiometer. A 50 $\Omega$ resistor serves as the internal noise source for this conversion. Its continuously monitored temperature facilitates the radiometer's switching between the antenna signal (radiometric power) and the noise source at a frequency of 3 Hz. The soil signal (radiometric power) is then divided by the calibration signal and multiplied by the physical temperature of the noise source to obtain the brightness temperature.

To determine the volumetric water content (VWC), an empirical relationship, which assumes a linear negative relationship between $\text{T}_{\text{sys}}$ and VWC is employed.  This equation is expressed as
\begin{equation}
	\text{VWC}=-a\text{T}_{\text{sys}} + b
	\label{eq:vwc_tsys}
\end{equation}
In Eq.~(\ref{eq:vwc_tsys}), $a > 0$, and its value is assumed constant during each crop growing season. The variable $b$  is adjusted for each pivot run,  and its value is influenced by changes in the crop canopy effect on the soil's radio emission. Parameters $a$ and $b$ are empirically determined by comparing the radiometer signal levels with measurements from a time-domain reflectometry soil moisture probe.

\subsection{Numerical Modeling}
The quadrant being studied has a total radius of 290 meters, a total depth of 0.6 meters, and a total angle of 0.5$\pi$ radians. To discretize the quadrant, the radius and angle were each divided into 30 and 17 equally spaced sectors, respectively. The number of nodes in the radial and azimuthal directions were chosen based on the distance between two consecutive microwave sensor measurements in the $r$ and $\theta$ directions. The depth $H_z$ was divided into 10 unequally spaced sectors with finer discretization near the surface and coarser discretization away from the surface. In total, the quadrant was discretized into $N_x = 5100$ states. Further mesh refinement in the $z$ direction was found to have a negligible impact on the state trajectories. Hence, the adopted spatial discretization ensured an accurate numerical approximation of the investigated field. A time step size of 10 minutes was used for the temporal discretization of the field model.

\subsection{Data Preparation and Pre-processing}
The study employed moisture content measurements taken from June 3rd, 2021 to July 22nd, 2021 in the investigated quadrant. Averagely, the microwave radiometers provided soil moisture measurements after every 30 seconds. 
Data assimilation was carried out using all the measurements obtained in June 3rd to June 29th, while for the remaining days in the simulation period, 80\% of the measurements were used for data assimilation and the remaining measurements were used for validation purposes. To ensure an appropriate representation of data for the state and parameter estimator, the raw measurements were taken through a series of data pre-processing steps, which are listed and explained below.
\begin{enumerate}
	\item \textbf{Sorting measurements by date and time}: Due to the large size of Field 3, the center pivot requires 2 to 3 days to complete its rotation cycle. Consequently, the water content measurements obtained from the microwave sensors at the end of the rotation cycle consist of measurements taken at various times over the course of 2 to 3 days. Therefore, it is important to sort the raw measurements initially by the date they were obtained. Then, all the measurements obtained on a particular day are arranged in ascending order according to their respective time.
	
	\item \textbf{Sorting measurements by quadrants}: As this study focuses on one specific quadrant of Field 3, it is necessary to sort the measurements collected on a particular day by the quadrant in which they are located.
	
	\item \textbf{Inferring the movement of the center pivot}: Since the microwave radiometers measure the soil water content of the field as the center pivot rotates, it is possible to infer the movement of the center pivot by analyzing how the measurement locations change over time. To accomplish this, the measurements are grouped based on a specific sampling time, denoted as $T_s$, so that the change in the measurement locations over time represents the circular movement of the pivot. Accurately determining the irrigated nodes of the quadrant at each sampling time requires inferring the movement of the center pivot. In this study, it was observed that grouping the measurements according to $T_s$ = 10 minutes accurately modeled the movement of the center pivot.

	\item \textbf{Dropping outliers}: Finally, the data set is processed to identify and remove extreme soil moisture content measurements. The saturated and residual soil moisture contents of the dominant soil type in the investigated quadrant are used to identify these extreme measurements. Any measurement exceeding the saturated moisture content or falling below the residual soil moisture content is excluded from the data set.
	
	\item \textbf{Mapping measurements to the nodes of the field model}: In order to associate each pre-processed measurement with a node in the field model, GPS coordinates are generated in a layout that matches the arrangement of nodes in the field model. When a measurement is received, the distances between its GPS coordinates and the generated coordinates are calculated, and the measurement is assigned to the node with the smallest distance from the measurement location.

\end{enumerate}
Weather data, including daily evapotranspiration (Figure~\ref{fig:et_rain_irrig_chapter_5}(\textbf{B}) in Appendix~\eqref{sec:weather_chapter_5}), daily average temperature (Figure~\ref{fig:et_rain_irrig_chapter_5}(\textbf{A}) in Appendix~\eqref{sec:weather_chapter_5}), and daily rainfall amounts (Figure~\ref{fig:et_rain_irrig_chapter_5}(\textbf{C}) in Appendix~\eqref{sec:weather_chapter_5}), were obtained from the Alberta Information Service website. These inputs are critical for the field model simulations. Similarly, irrigation application depths for the simulation period (Figure~\ref{fig:et_rain_irrig_chapter_5}(\textbf{C}) in Appendix~\eqref{sec:weather_chapter_5}) were sourced from the Research Farm. Additionally, the crop coefficient for the simulation period was determined using Equation~\eqref{eq:kc_relation_barley}, provided in Appendix~\eqref{sec:kc_relations}.

\subsection{Sensitivity Analysis and Orthogonal Projection}
The sensitivity matrices for each sector of the investigated quadrant were generated using the numerical model and nominal hydraulic parameters illustrated in Figure~\ref{fig:soil_pars_sens_analyis} in Appendix~\eqref{sec:hyd_pars_sens_analysis}. These matrices were created for all 2,550 hydraulic parameters associated with the nodes located within the sensing depth of the microwave radiometers. After applying the orthogonal projection method to the rank-deficient, scaled sensitivity matrices, it was determined that the most estimable parameters were the 5 hydraulic parameters associated with the measured locations on the field. The insights obtained from these results were used to determine the most estimable parameters at each time step during the real case study.

In this study, for the measurement sampling time of $T_s=10$ minutes, it was observed that there are between 10 and 60 measurements taken at each sampling time. By applying the results of the orthogonal projection, between 50 and 300 parameters can be uniquely estimated at each sampling time, together with the 5,100 states that make up the field model. 

{Considering the 20 cm sensing depth of the microwave radiometers and the outcomes of the parameter selection process, the spatial distribution of these nodal hydraulic parameters is replicated across all axial nodes within the radiometers' sensing depth.  For spatial nodes beyond the penetration depth of the microwave radiometers, where their nodal parameters have minimal impact on the observations under investigation, nominal values corresponding to the hydraulic parameters of sandy clay loam, the predominant soil type in the studied quadrant, were assigned.} At each time step of the estimation, the values of the non-estimable parameters are updated using the Kriging interpolation method once the estimated parameters for the measured nodes are obtained to enhance the accuracy of soil moisture estimation. Furthermore, as the proposed approach aims to estimate the entire soil moisture distribution across the investigated field, it facilitates the estimation of the complete soil moisture content within the 0.6 m field, which is represented by the 5100 states in the actual case study.

\subsection{State and Parameter Estimator Design}
The estimator was initialized with a covariance matrix, as described in Section~\ref{sec:State and Parameter Estimator}, that accounted for all potentially estimable parameters, amounting to 2,550 parameters. The EKF was initialized with an initial estimate $\hat{x}_a(0)$ and a diagonal covariance matrix $P_a(0|0)$, both set based on the limited information available about the initial state $x_a(0)$. To reflect the high uncertainty in $\hat{x}_a(0)$, the diagonal entries of $P_a(0|0)$ were assigned large values: 340 for the state variables and 6 for the parameters. The process noise covariance matrix $Q$ and the measurement noise covariance matrix $R$ were treated as tuning parameters. Their diagonal elements were gradually increased until a satisfactory agreement was achieved between the estimated and observed soil moisture values from the EKF filtering step. In the final configuration, $Q$ was set to $10I_{7,650}$, and $R$ was set to $0.01I_{N_y}$, where $I_{7,650}$ is the identity matrix of size 7,650, and $N_y$ represents the number of measurements collected at each sampling time.

\subsection{Evaluation Criteria}
The performance of the proposed soil moisture and hydraulic parameter estimation approach was assessed with two types of cross-validation. In the first type, the measurements acquired at each sampling time were randomly split into a training set and a validation set. The estimates provided by the proposed approach  were compared with the measurements in the validation set. Essentially, this assessment validated the accuracy of the soil moisture estimates.

The second type of cross-validation involved using all measurements obtained on a specific day within the simulation period, precisely on July 21st, 2021, for validation purposes. This process involved simulating the field model, considering the applied irrigation and the observed weather conditions on July 21st, 2021. To perform this validation, a spatial map comprising all observations obtained from the microwave sensors on July 21st, 2021, was generated for the investigated quadrant. The Kriging interpolation method was utilized to create this spatial map. To validate the predictions from the field model,  a spatial map showing the average soil water content predictions for the top 20 cm of the field model was generated. Subsequently, absolute errors between the actual moisture observations and the predicted soil moisture contents were calculated. These errors were spatially plotted to enable a quantitative comparison between the actual and predicted soil moisture contents. Essentially, this validation assessed the predictive capacity of the field model after the estimated states and parameters have converged.

The normalized root mean square error (NRMSE) was used to quantitatively assess the performance of the proposed approach.
The NRMSE was calculated for the cross-validation step by comparing the estimated soil moisture content ($\hat{y}_k$) with the measured soil moisture content ($y_k$) values. The NRMSE is defined as:
\begin{equation}
	\text{NRMSE} = \frac{1}{(y_{\text{max}} - y_{\text{min}})}\sqrt{\frac{\sum_{k=1}^{k=N}(y_k - \hat{y}_k)^2}{N}}	
\end{equation}
where $N$ denotes the total number of measurements in the validation set, and $y_{\text{max}}$ and $y_{\text{min}}$ represent the maximum and minimum soil moisture content values in the validation set, respectively. A smaller value of NRMSE indicates a better match between the estimated and measured values.

To further highlight the advantages of estimating soil moisture and hydraulic parameters simultaneously, the results of two additional case studies were presented and analyzed together with the results of the proposed approach. The first case study (Case Study 1) involved estimating soil moisture using the hydraulic parameters of the dominant soil type in the investigated field, specifically sandy clay loam. The hydraulic properties of sandy clay loam are shown in Table~\ref{tbl:hydr_pars_scl}. The second case study (Case Study 2) employed hydraulic parameters obtained from a soil texture survey conducted in the investigated field, where the parameters were interpolated at the several sampling locations using the Kriging method. The spatial distributions of the interpolated hydraulic parameters are depicted in Figure~\ref{fig:spat_pars_texture_survey} in Appendix~\eqref{sec:hyd_pars_texture_survey}. It is worth noting that the hydraulic parameters shown in these figures were obtained using PTFs proposed in~\cite{saxton1986estimating,carsel1988developing}, which took into account the percentages of sand, silt, and clay determined from the soil texture survey.
\begin{table}[t]
	\caption{The hydraulic parameters of sandy clay loam soil~\cite{carsel1988developing}.}
	\small 
	\centering
	\begin{tabular}{ccccc}
		\hline
		\hline
		{$K_{s}$ (m/s)} &{$\theta_{s}$ $(\text{m}^{3}/\text{m}^{3})$}&{$\theta_{r}$ $(\text{m}^{3}/\text{m}^{3})$}& {$\alpha$ (1/m)}& {$n$ (-)}\\
		\hline
		$7.222\times 10^{-7}$  & 0.410 & 0.090&1.90&1.31\\
		
		\hline
	\end{tabular} \vspace{4mm} \label{tbl:hydr_pars_scl} 
\end{table}

\subsection{Results and Discussion}

\subsubsection{Estimation Performance}

\begin{figure}[!ht]
	\centering
	\subfigure[Case 1]{
		\includegraphics[width=0.4\textwidth]{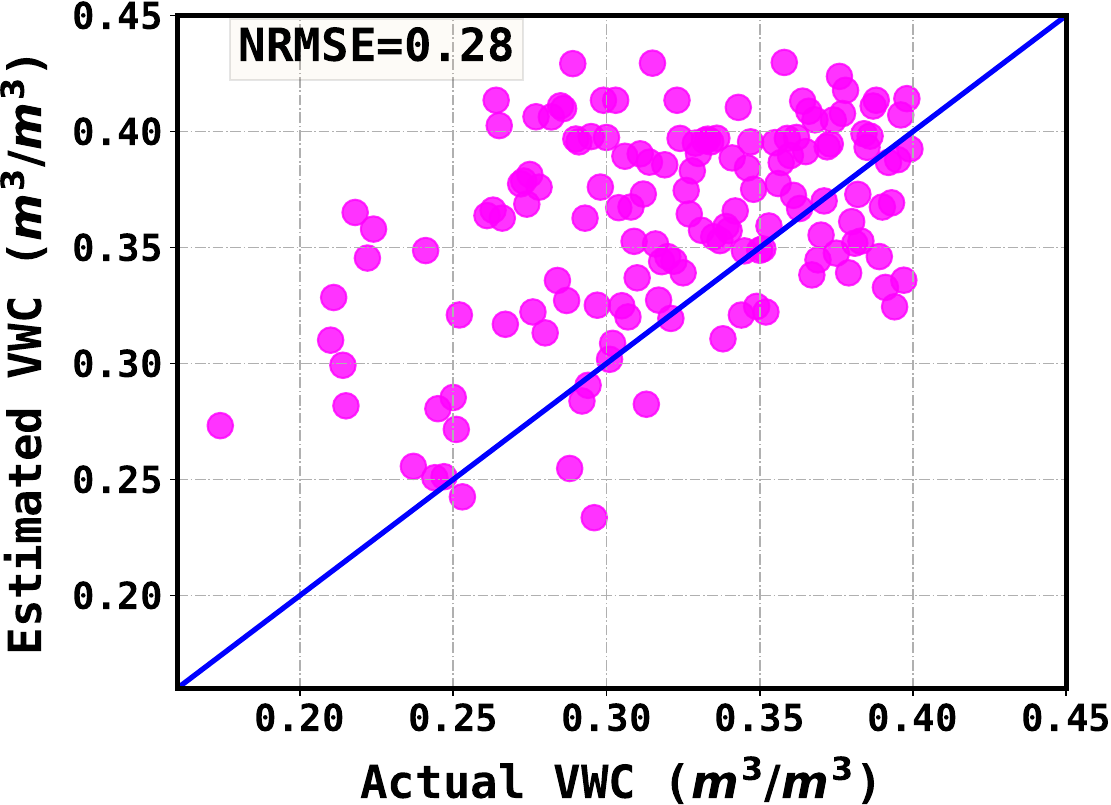}}%
		
	\subfigure[Case 2]{
		\includegraphics[width=0.4\textwidth]{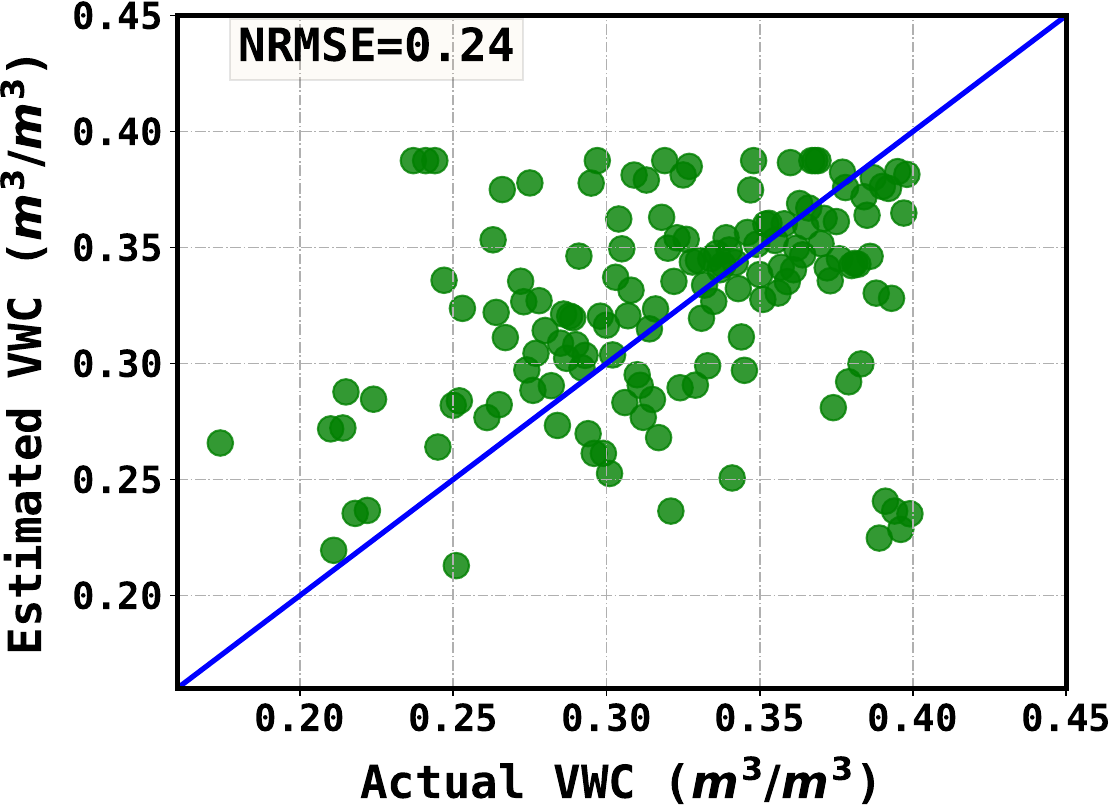}}
		
	\centering
	\subfigure[Case 3]{
		\includegraphics[width=0.4\textwidth]{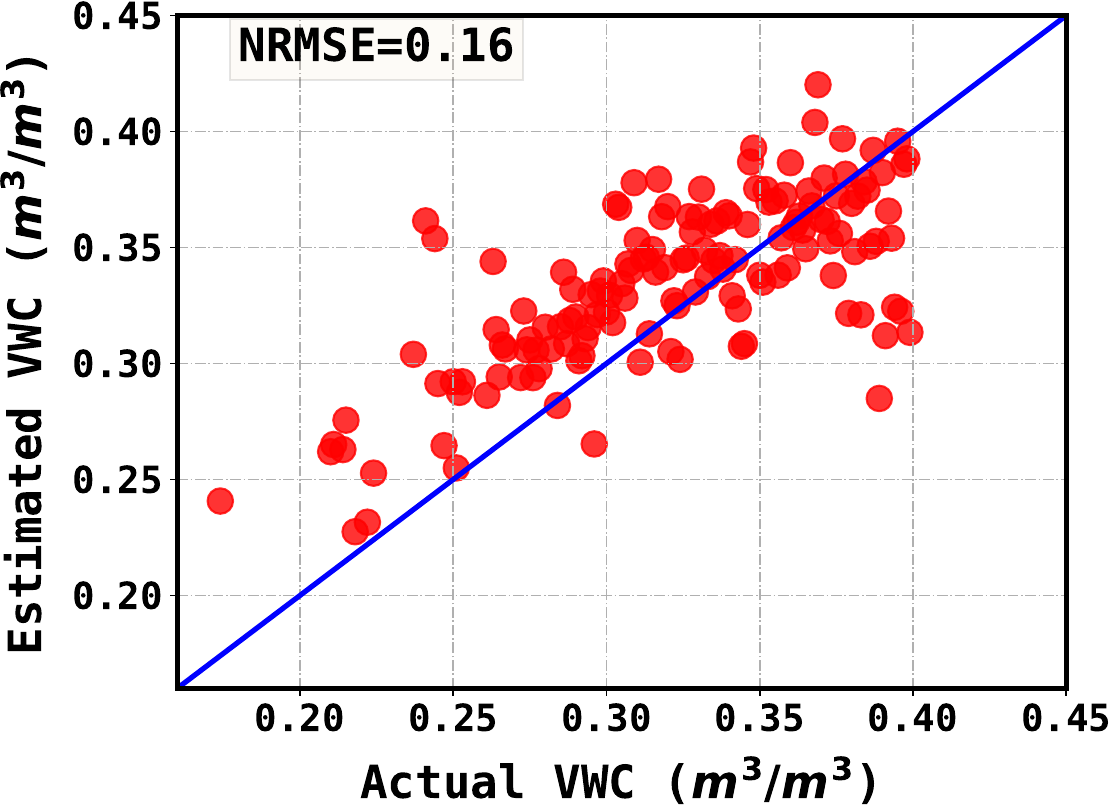}}
	\caption{Cross-validation results for July 2nd, 2021.}
	\label{fig:valid_02}
\end{figure}

\begin{figure}[!ht]
	\centering
	\subfigure[Case 1]{
		\includegraphics[width=0.4\textwidth]{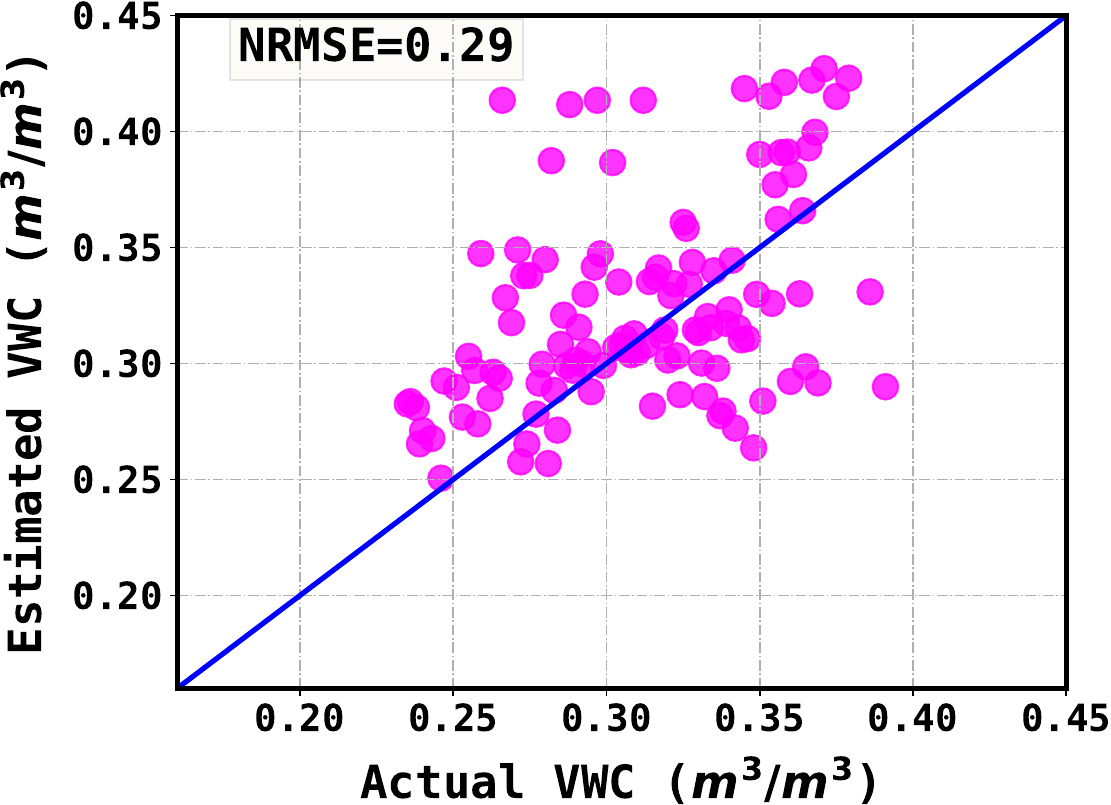}}%
		
	\subfigure[Case 2]{
		\includegraphics[width=0.4\textwidth]{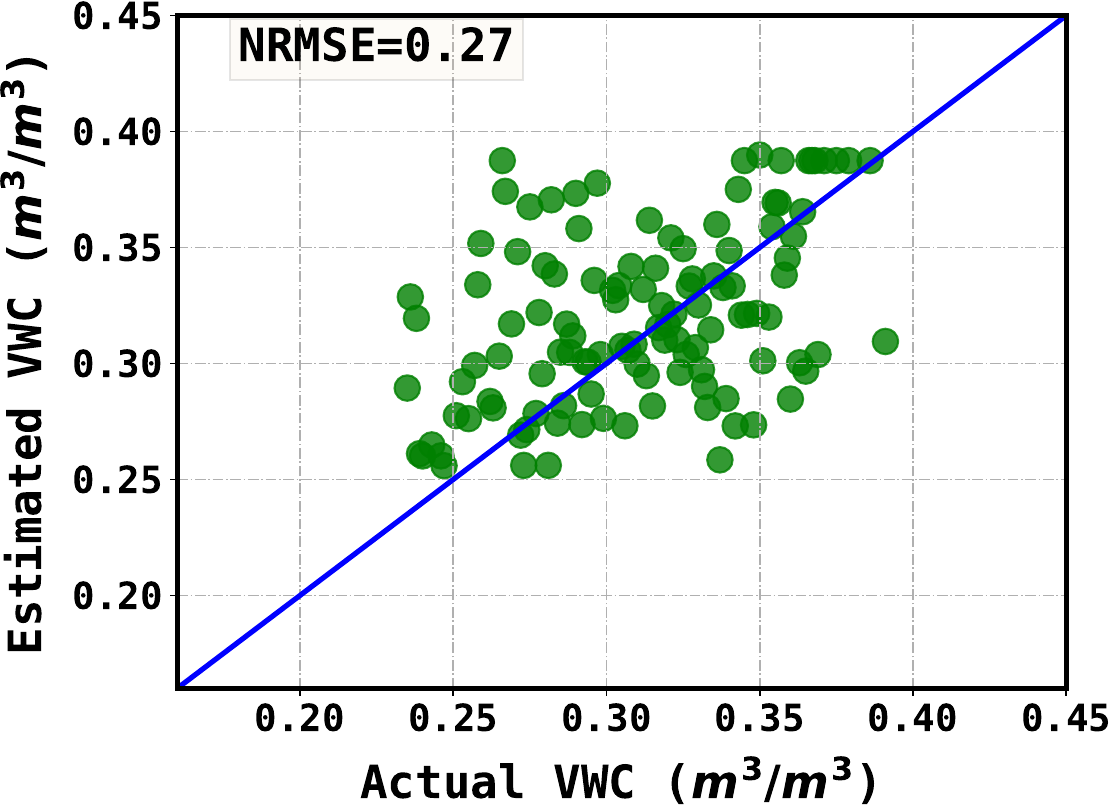}}
		
	\centering
	\subfigure[Case 3]{
		\includegraphics[width=0.4\textwidth]{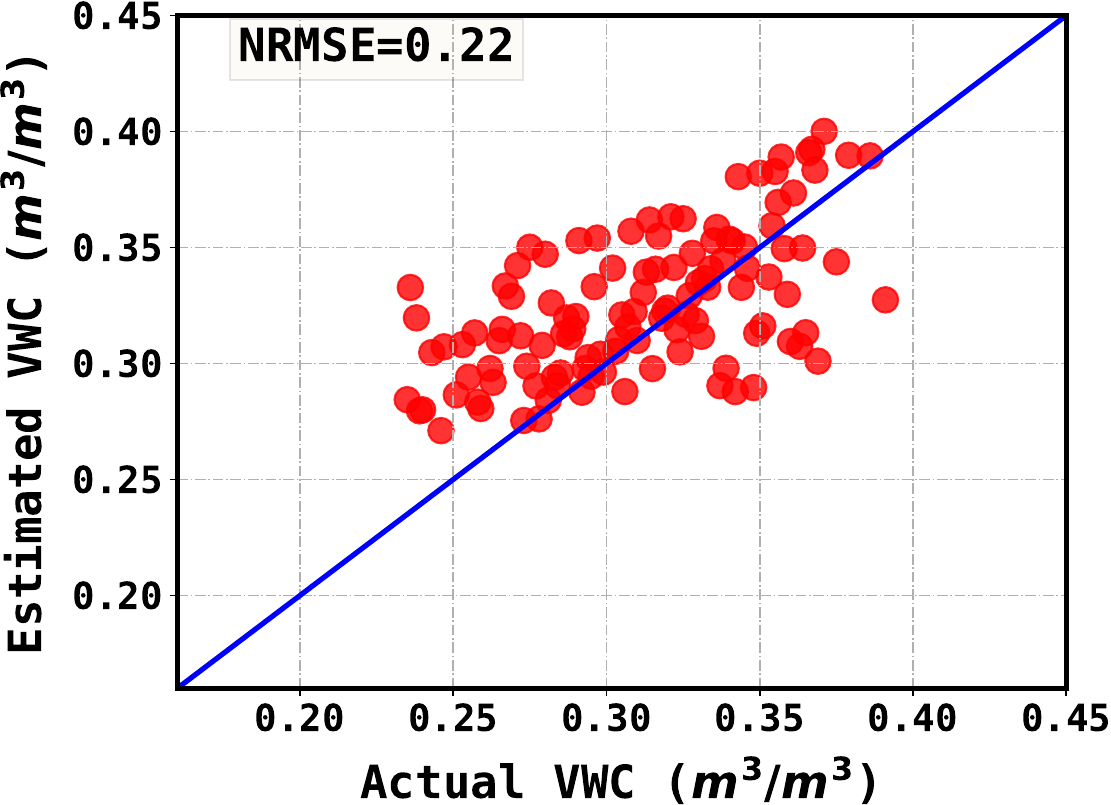}}
	\caption{Cross-validation results for July 5th, 2021.}
	\label{fig:valid_05}
\end{figure}

The results of the first type of cross-validation performed on July 2nd and July 5th are depicted in Figures \ref{fig:valid_02} and~\ref{fig:valid_05}. The inclusion of the $y=x$ line across all three case studies allows for a comparative evaluation of their respective performances. Overall, these figures validate the conclusions drawn in prior studies~\cite{agyeman2021soil,medina2014kalmanp3,montzka2011hydraulic}, which effectively used real-time near-surface soil moisture observations from remote sensors to estimate both soil moisture and hydraulic parameters based on the Richards equation. A comprehensive analysis of these figures, combined with the computed NRMSE for the three Cases, demonstrates that the proposed approach produces the most accurate estimates, followed by Case 2, and then Case 1. This trend is evident from the validation points clustering more tightly around the $y=x$ line for Case 3, followed by Case 2, and then Case 1. By using the soil moisture estimates in Case 1 as a reference point, it is observed that, for the validation results on July 2nd and 5th, the proposed approach enhances the accuracy of soil moisture estimates by 43\% and 24\%, respectively. The estimation results of Case 1 highlight the decline in estimation performance associated with using hydraulic parameters of the dominant soil type in the studied field for soil moisture estimation, as performed in~\cite{agyeman2021soil}. This approach notably diminishes the accuracy of soil moisture estimates since a single hydraulic parameter inadequately captures the inherent variability in soil texture across the field. The enhanced soil moisture estimation performance in Cases 2 and 3 compared to Case 1 aligns with findings reported in~\cite{medina2014kalman}, indicating that incorrectly identified hydraulic parameters introduce biases and errors in soil moisture estimates. Moreover, the reduced estimation performance in Case 2 compared to Case 3 also aligns with conclusions drawn from~\cite{li2007estimating,soet2003functional}, suggesting that relying solely on pedo-transfer functions for determining soil hydraulic parameters, as performed in Case 2, results in modeling errors.

Additionally, Figures~\ref{fig:valid_02} and~\ref{fig:valid_05} depict noticeable uncertainty in soil moisture estimates, evident from the dispersion of validation data around the $y=x$ lines. This uncertainty is more prominent in Cases 1 and 2 than in Case 3. Case 1's sole reliance on hydraulic parameters of the dominant soil type leads to increased uncertainty due to limited representation of soil variability. Case 2 exhibits considerable uncertainty attributed to soil moisture variability beyond what soil texture accounts for. In contrast, Case 3, involving a careful estimation of a subset of hydraulic parameters alongside soil moisture, displays reduced uncertainty.  The residual uncertainty in the estimates of the proposed approach arises from several sources. Notably, estimating only a systematically selected subset of the total hydraulic parameters in the field model contributes to the uncertainty in soil moisture estimates. While the non-estimable parameters minimally affect the accuracy of the field model, their unknown precise values can increase uncertainties in soil moisture estimates. It has been reported in~\cite{chen2022parameter} that if the system under study is nonlinear in the parameters, the nominal values assigned to the non-estimable parameters can affect the prediction accuracy of the model. Considering the nonlinear relationship between the Richards equation and parameters, the nominal values of non-estimable parameters could potentially impact model accuracy. Furthermore, noise present in soil moisture observations, along with discrepancies between the field model and the actual field (referred to as process disturbance), can introduce additional uncertainties in the estimates. Although the estimation algorithm quantifies the measurement noise and the process disturbance, some aspects of these factors remain unaccounted for, leading to uncertainties in soil moisture estimates. Exploring methods to further reduce uncertainty could be a subject for future study. For instance, increasing the number of estimable hydraulic parameters by combining remotely sensed soil moisture observations with point sensor measurements might offer a potential solution.
\begin{figure}[!ht]
	\centering
	\subfigure[Actual.]{
		\includegraphics[width=0.4\textwidth]{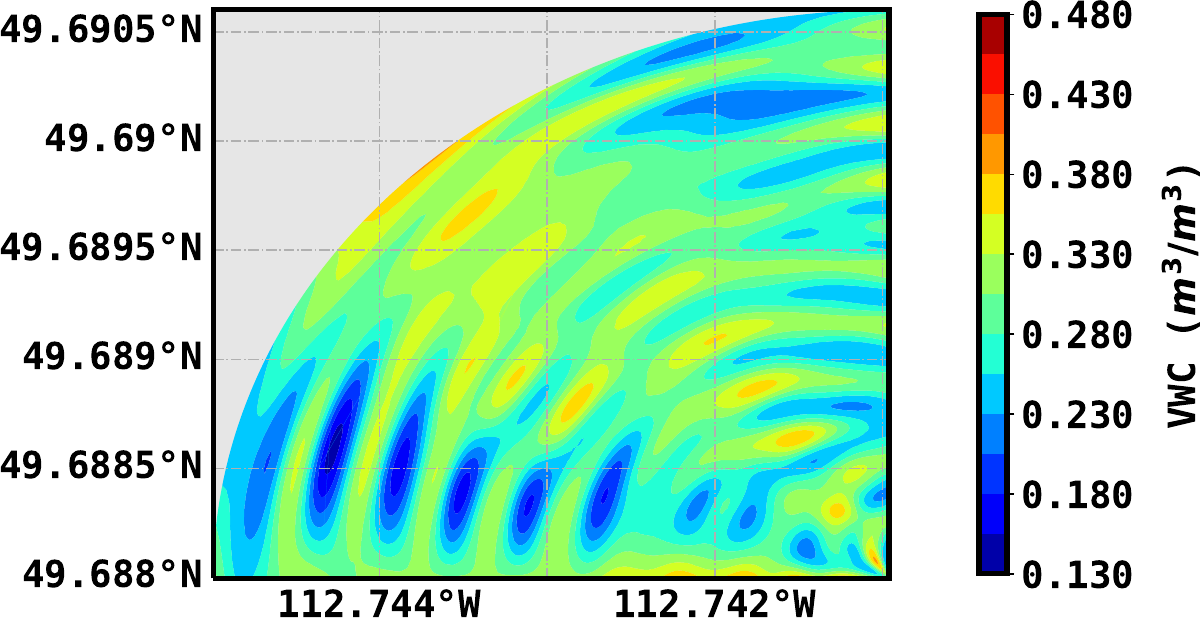}}%
		
	\subfigure[Estimated.]{
		\includegraphics[width=0.4\textwidth]{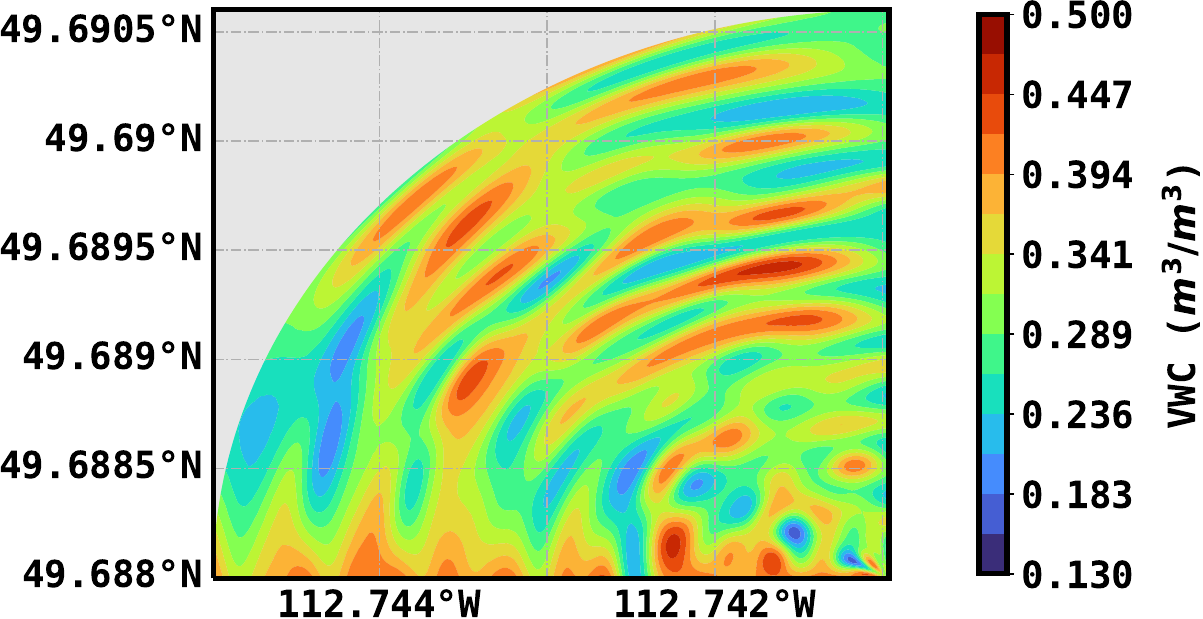}}
		
	\centering
	\subfigure[Absolute error.]{
		\includegraphics[width=0.4\textwidth]{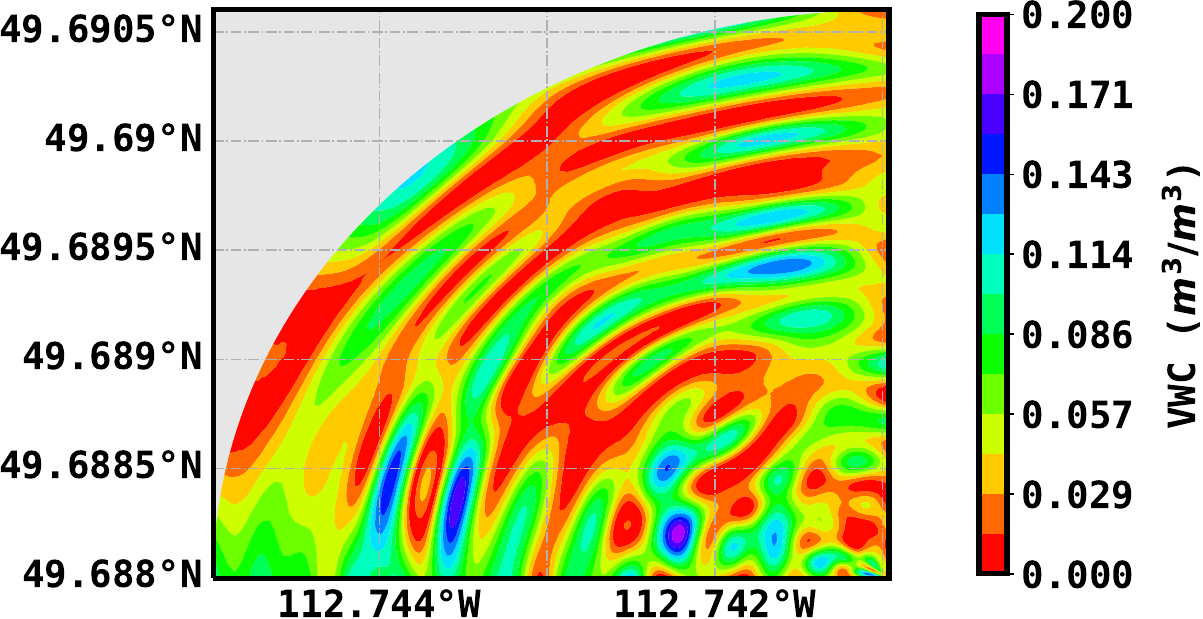}}
	\caption{Cross validation using all the measurements obtained on July 21st, 2021 under Case 2.}
	\label{fig:valid_2_se}
\end{figure}
\begin{figure}[!ht]
	\centering
	\subfigure[Actual.]{
		\includegraphics[width=0.4\textwidth]{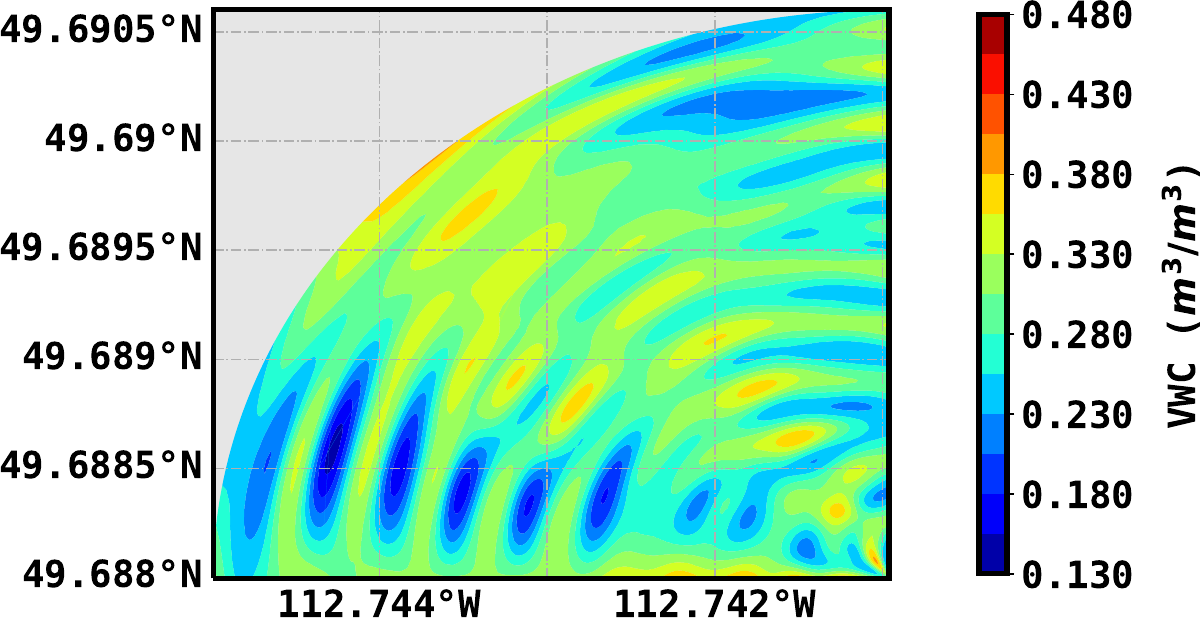}}%
		
	\subfigure[Estimated.]{
		\includegraphics[width=0.4\textwidth]{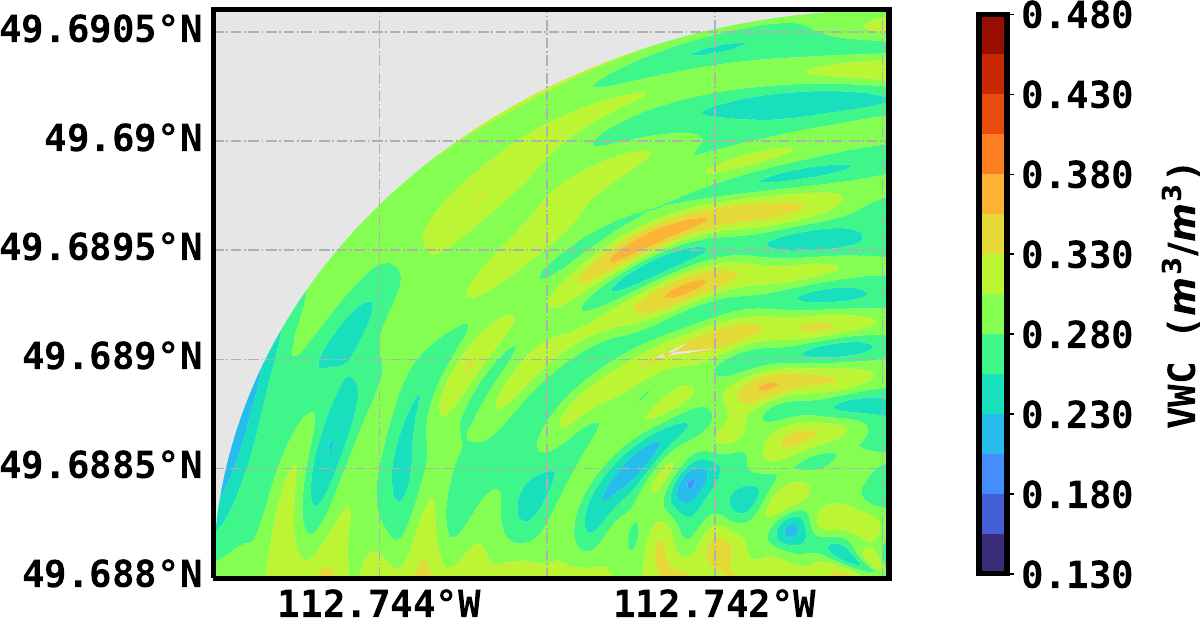}}
		
	\centering
	\subfigure[Absolute error.]{
		\includegraphics[width=0.4\textwidth]{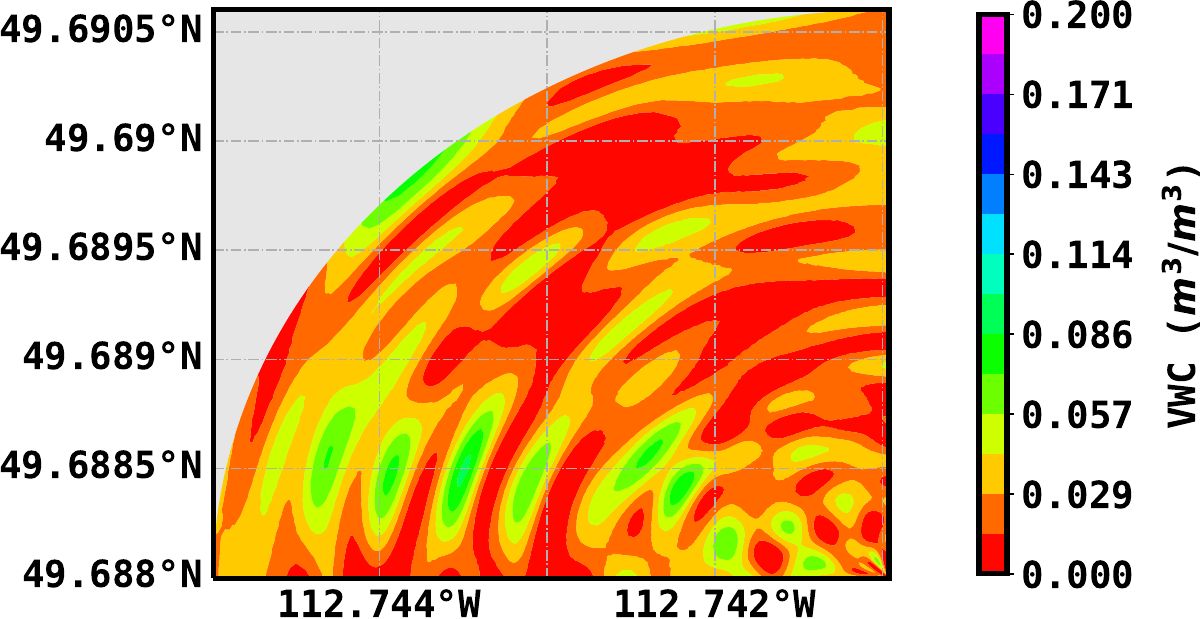}}
	\caption{Cross validation using all the measurements obtained on July 21st, 2021 under Case 3.}
	\label{fig:valid_2_sep}
\end{figure}
\subsubsection{Predictive Performance}
In the second type of cross validation, intended to evaluate the precision of the proposed approach in short-term soil moisture predictions, the least accurate outcome derived from state estimation using dominant soil type parameters is excluded, given its prior inadequate performance in the initial cross-validation. The validation outcomes for Case Study 2 and the proposed approach are illustrated in Figures \ref{fig:valid_2_se} and \ref{fig:valid_2_sep}, respectively, while the corresponding NRMSE values are summarized in Table \ref{tbl:perf_july21}. These visual representations notably depict substantial variability in soil moisture, primarily stemming from the inherent spatial heterogeneity within the field. This spatial heterogeneity leads to varying soil moisture levels across the field, even under uniform irrigation, owing to diverse water-holding capacities among different soil types. Overall, these findings align with conclusions outlined in~\cite{agyeman2021soil}, suggesting that estimating soil moisture using near-surface soil moisture measurements from remote sensors yields a field model capable of accurately predicting short-term soil moisture levels. This holds particular significance in instances where remote sensors cannot provide measurements, but soil moisture information is critical for determining irrigation application depths. Figures~\ref{fig:valid_2_se}(c) and~\ref{fig:valid_2_sep}(c) indicate that the proposed approach offers the most precise model prediction, with a maximum absolute error of approximately 0.086 $\text{m}^3/\text{m}^3$, compared to the second case study's maximum absolute error of about 0.20 $\text{m}^3/\text{m}^3$. This observation is reinforced by the NRMSE values summarized in Table~\ref{tbl:perf_july21}, where the proposed approach yielded the smallest NRMSE value. Using the results obtained in Case Study 2 as a benchmark, it is evident that the proposed approach enhances the predictive capability of the field model by 50\%. Notably, the NRMSE of the proposed approach is smaller in the second type of cross-validation compared to the first type due to considering more samples in the latter. Generally, after the convergence of states and parameters, the estimation accuracy is expected to exceed the predictive accuracy of the field model.

\subsubsection{Estimated Hydraulic Parameters}
\begin{figure}[!ht]
	\centering
	\subfigure[Estimates.]{
		\includegraphics[width=0.45\textwidth]{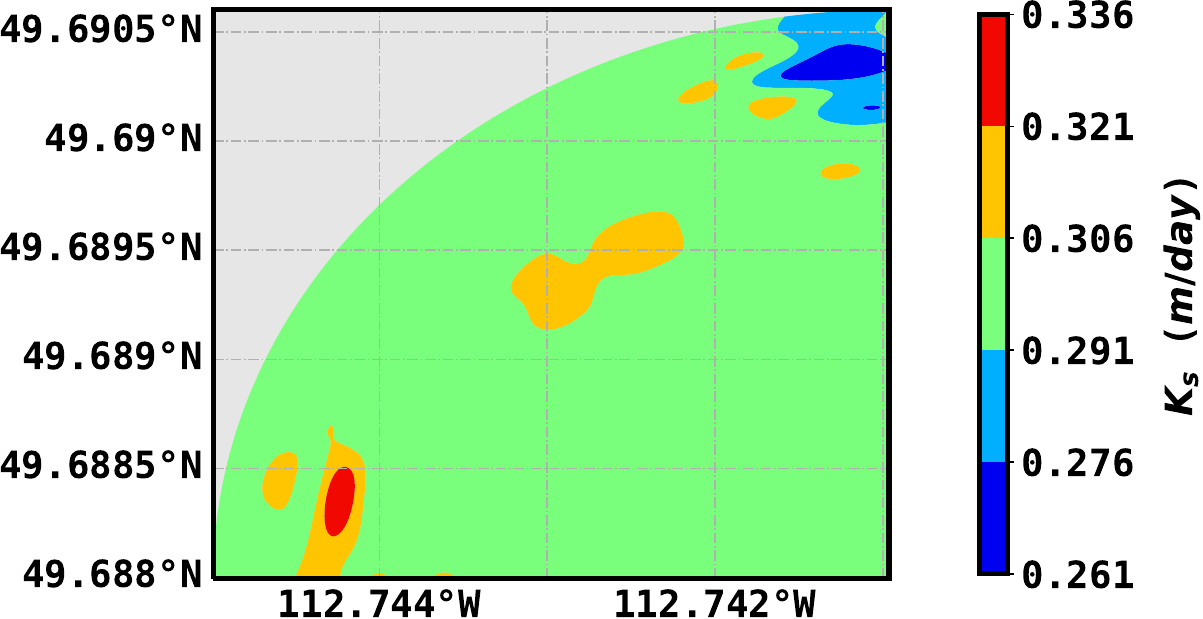}}%
		
	\subfigure[Uncertainty.]{
		\includegraphics[width=0.45\textwidth]{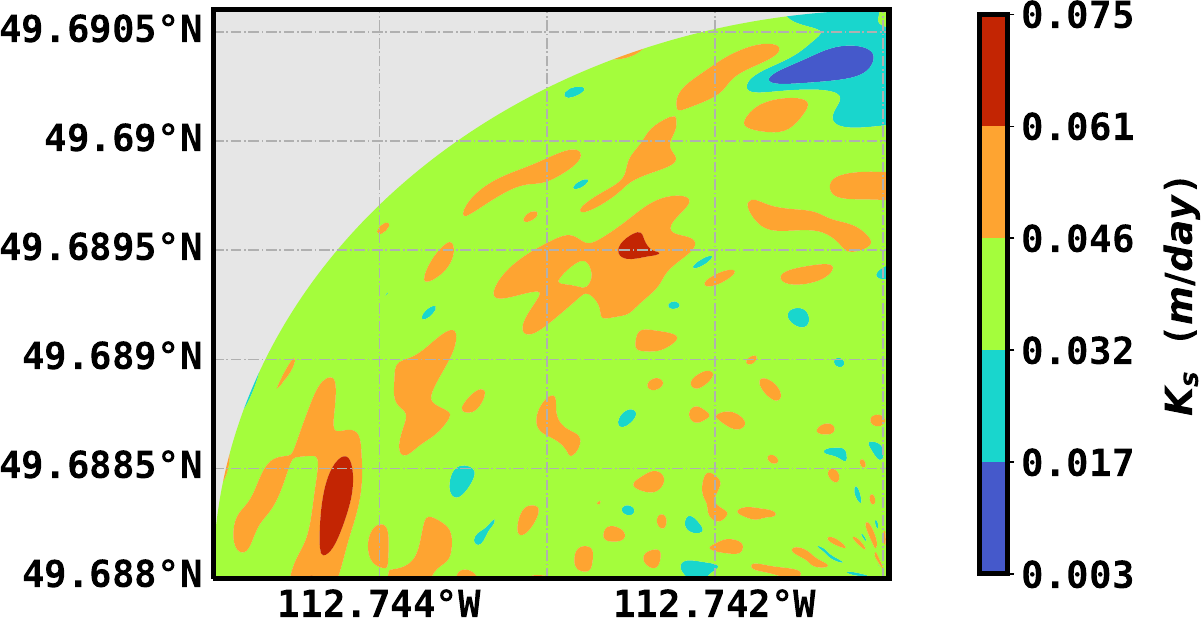}}
	\caption{{Estimated $K_s$ (m/day) at the end of the estimation period.}}
	\label{fig:estimated_ks}
\end{figure}

\begin{figure}[!ht]
	\centering
	\subfigure[Estimates.]{
		\includegraphics[width=0.45\textwidth]{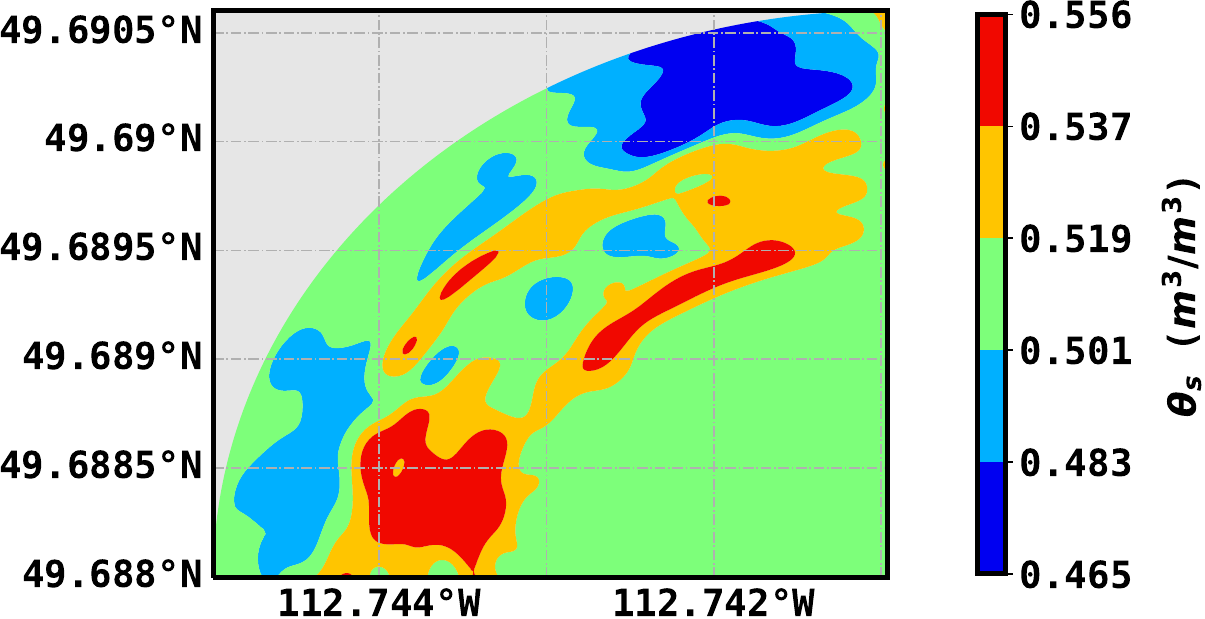}}%
		
	\subfigure[Uncertainty.]{
		\includegraphics[width=0.45\textwidth]{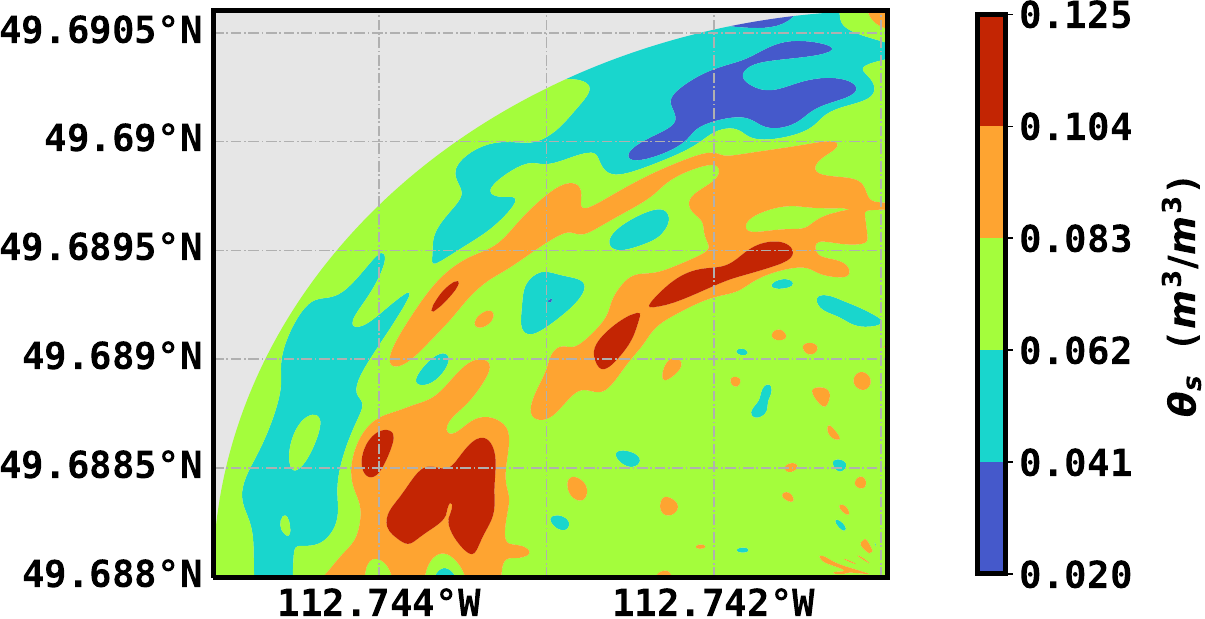}}
	\caption{{Estimated $\theta_s$ ($m^3/m^3$) at the end of the estimation period.}}
	\label{fig:estimated_ts}
\end{figure}

The spatial map as well as the uncertainty map of the estimated hydraulic parameters are depicted in Figures~\ref{fig:estimated_ks} and \ref{fig:estimated_ts}.  It is noteworthy that the estimated parameters correspond to the top 20 cm of the investigated field.
Furthermore, this study evaluated the accuracy of the hydraulic parameter estimates, focusing specifically on the $K_s$. As depicted in Figure~\ref{fig:estimated_ks}, the estimated $K_s$ converged to values ranging from 0.261 to 0.336 $\text{m}/\text{day}$. To validate the reliability of the estimated $K_s$, laboratory-determined $K_s$ values for five randomly sampled points were compared with the estimated $K_s$ values. A good agreement was found, with the mean value being 0.39 $\text{m}/\text{day}$ for these laboratory-determined $K_s$ values. These results indicate that the proposed approach converges to hydraulic conductivity values that are representative of the entire investigated quadrant. These findings are consistent with previous studies, such as those reported in~\cite{medina2014kalmanp3}, where assimilating near-surface soil moisture observations into the Richards equation resulted in reasonably accurate $K_s$ values. Future studies will be necessary to extend this validation approach to other parameters, whenever their actual values are available, to ensure a comprehensive evaluation of the proposed method's performance in estimating various soil hydraulic parameters.

{Figures~\ref{fig:estimated_ks} and ~\ref{fig:estimated_ts} also demonstrate the estimation approach's ability to capture spatial variability in hydraulic parameters, while indicating its inability to capture variations in soil hydraulic parameters along the axial (vertical) direction. Instead, the estimation approach yields a hydraulic parameter set that reflects the sensing depth of the microwave radiometers. These findings can be explained by considering insights derived from the parameter selection step and the nature of the observations utilized in this study. It is noteworthy that microwave sensors capture spatial variability in soil moisture, providing the average soil water content in the upper 20 cm of the soil column along the axial (vertical) direction. Consequently, guided by the parameter selection process where the estimable parameters correspond to nodal hydraulic parameters of measured nodes, the parameter estimation results in a hydraulic parameter set representative of the upper 20 cm of the field. Essentially, with only a single measurement available in the axial direction for each measured location and considering insights from the parameter selection step, the estimation approach is unable to capture vertical heterogeneity in hydraulic parameters. Nonetheless, it effectively captures variability in soil hydraulic parameters within the radial-azimuthal plane (2D-plane). This observation highlights the limitation of a single sensing technique in adequately capturing the spatial and vertical variability in the hydraulic parameters of agro-hydrological systems. To address the limitation of microwave sensing in capturing vertical variability in soil hydraulic parameters, incorporating soil moisture measurements from point sensors in the proposed estimation framework could be considered, as point sensors have the capability to provide a vertical characterization of soil moisture content.}

Furthermore, unlike previous studies~\cite{leger2020evaluating,abbasi2012estimating} that utilized separate hydraulic parameter sets for dry and wet conditions to address soil hysteresis effects, this study employed a single parameter set applicable to all soil conditions. Through a recursive estimation approach driven by real-time soil moisture observations from the field, the method allowed hydraulic parameters to adapt to changing soil conditions. This adaptability enabled the  proposed approach to handle the hysteresis behavior observed in soils. However, further research is needed to compare this approach with alternative methodologies that specifically use separate parameter sets for dry and wet conditions to handle soil hysteresis effects.
\begin{table}[t]
	\caption{NRMSEs of the Type 2 cross-validation performed on July 21st, 2021.}
	\small 
	\centering
	\begin{tabular}{cc}
		\hline
		\hline
		\textbf{Case Study}& \textbf{NRMSE}\\
		\hline
		State estimation  with texture survey parameters & 0.24\\
		\hline 
		State and parameter estimation & 0.12\\
		\hline
	\end{tabular}\vspace{4mm}\label{tbl:perf_july21}
\end{table}
\section{Conclusion}
\label{ch3:sec:Conclusions}
In this paper, a systematic method to simultaneously estimate soil moisture and soil hydraulic parameters in large-scale agro-hydrological systems by utilizing soil moisture observations acquired through microwave radiometers installed on center pivot irrigation systems was proposed. Basically, the proposed approach involves modeling the field under study with the cylindrical coordinate version of the Richards equation, employing sensitivity analysis and the orthogonal projection methods to address issues of parameter estimability, and assimilating the remotely sensed moisture observations into the field model using the extended Kalman filtering technique. Technical issues, such as constructing the output sensitivity matrix to handle spatially varying measurements and modifying the extended Kalman filter to accommodate changing estimable parameters over time, were addressed.

The outcomes of sensitivity analysis and orthogonal projection methods show that the hydraulic parameters of the measured nodes in the field model are the estimable parameters, and these parameters, along with all states of the field model, can be reliably and uniquely estimated.  Simulated and real case studies were conducted, revealing that the proposed approach enhances the accuracy of soil moisture estimation while providing reliable estimates of hydraulic parameters. In summary, the results from the proposed approach can serve as a good basis for the calibration of agro-hydrological models for the purposes of closed-loop irrigation scheduling.

\section{Acknowledgements}
Financial support from Natural Sciences and Engineering Research
Council of Canada and Alberta Innovates is gratefully acknowledged.
\appendix
\section*{Appendix}
\section{Crop Coefficient of Barley}\label{sec:kc_relations}
\begin{multline}
	\label{eq:kc_relation_barley}
	K_c(g) = 0.04217 + 0.001508g + \left(4.89\times10^{-6}\right)g^2 -  \\ \left(8.69\times10^{-9}\right)g^3 + \left(2.49\times10^{-12}\right)g^4
\end{multline}
\normalsize
where $g$ is the cumulative growing-degree days (GDD). GDD is calculated as follows:
\begin{equation}
	\text{GDD} = \text{T}_{\text{avg}} - \text{T}_{\text{base}}
\end{equation}
where $\text{T}_{\text{avg}}$ is the daily average/mean temperature and $\text{T}_{\text{base}}$ is the base temperature below which crop growth ceases (5\textdegree C). Equation~\eqref{eq:kc_relation_barley} was obtained from from~\citep{bennett2011crop}.

\section{Spatial Map of Nominal Soil Hydraulic Parameters}\label{sec:hyd_pars_sens_analysis}
\begin{figure}
	\centering
	\subfigure[${\theta_s~(m^3m^{-3})}$]{
		\includegraphics[width=0.2\textwidth]{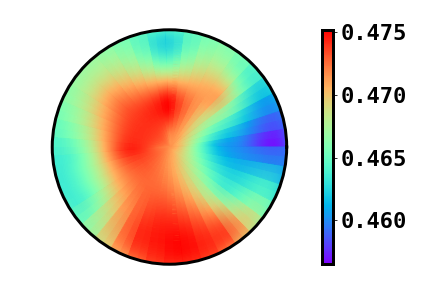}
		\label{fig:sub1}} 
	\subfigure[${\theta_r~(m^3m^{-3})}$]{
		\includegraphics[width=0.2\textwidth]{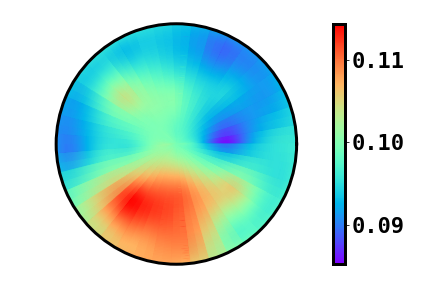}
		\label{fig:sub2}}
	\subfigure[${K_{s}~(m/day)}$]{
		\includegraphics[width=0.2\textwidth]{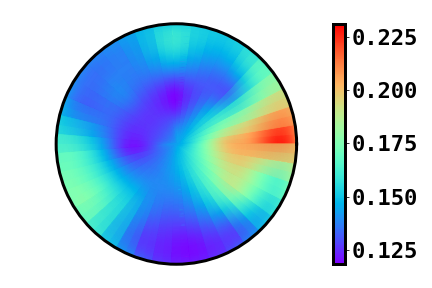}
		\label{fig:sub3}}
	\centering
	\subfigure[${\alpha~(m^{-1})}$]{
		\includegraphics[width=0.2\textwidth]{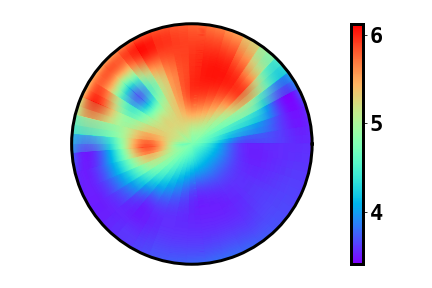}
		\label{fig:sub4}} 
	\subfigure[${n~(-)}$]{
		\includegraphics[width=0.2\textwidth]{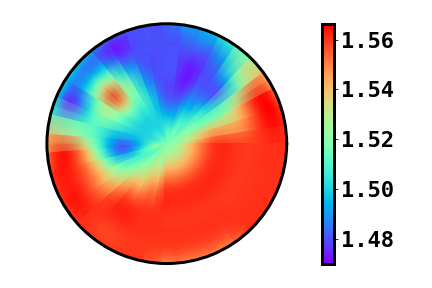}
		\label{fig:sub5}}
	\caption{Spatial distribution of the nominal soil hydraulic parameters.}
	\label{fig:soil_pars_sens_analyis}
\end{figure}

These parameters were derived from a soil texture survey conducted at a Research Farm operated by Lethbridge Polytechnic in Alberta, Canada. The survey sampled 60 locations at a depth of 30 cm and determined the clay, silt, and sand percentages at each site through laboratory analysis. Pedotransfer functions proposed by~\citep{saxton1986estimating,carsel1988developing} were utilized to calculate the 5 hydraulic parameters at these sites. The parameters were then spatially interpolated across the field using Kriging interpolation.

\section{Spatial Maps of Hydraulic Parameters From Texture Survey}\label{sec:hyd_pars_texture_survey}
\begin{figure}
	\centering
	\subfigure[Interpolated $\theta_s~(m^3m^{-3})$]{
		\includegraphics[width=0.2\textwidth]{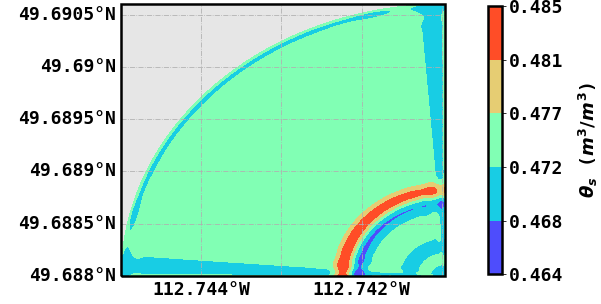}
	} %
	\subfigure[Interpolated $\theta_r~(m^3m^{-3})$]{
		\includegraphics[width=0.2\textwidth]{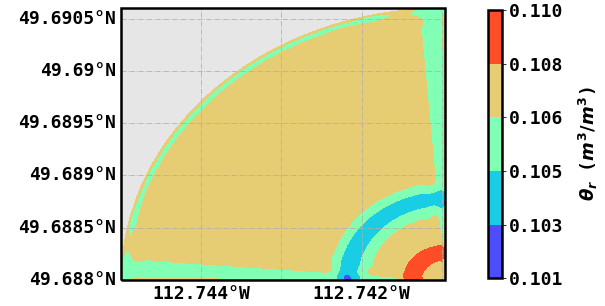}
	}
	\subfigure[Interpolated $K_{s}~(m {day}^{-1})$]{
		\includegraphics[width=0.2\textwidth]{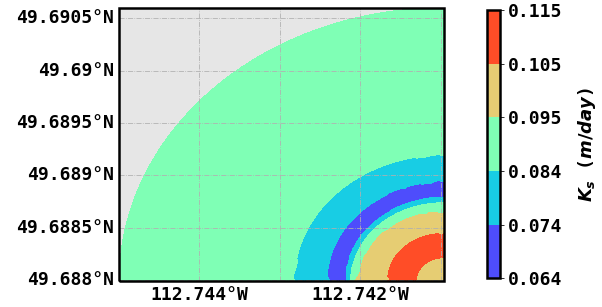}
	}
	\centering
	\subfigure[Interpolated $\alpha~(m^{-1})$]{
		\includegraphics[width=0.2\textwidth]{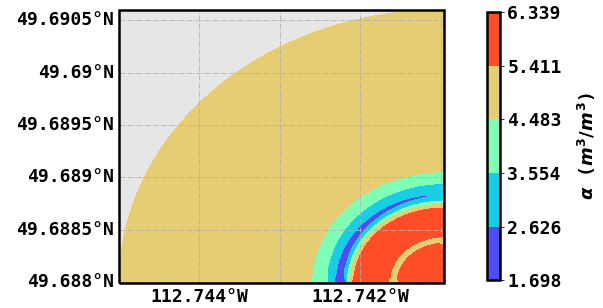}
	} %
	\subfigure[Interpolated $n~(-)$]{
		\includegraphics[width=0.2\textwidth]{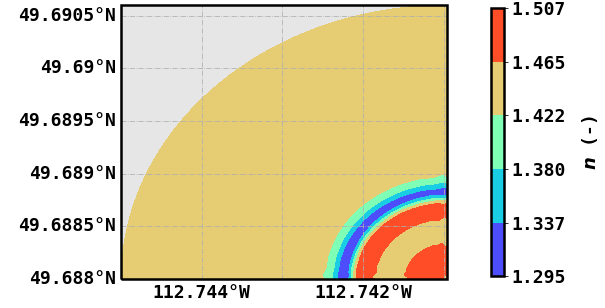}
	}
	\caption{Spatially interpolated hydraulic parameters derived from field texture survey.}
	\label{fig:spat_pars_texture_survey}
\end{figure}

\section{Reference Evapotranspiration, Rain, and Irrigation Amounts for Field Experiments}\label{sec:weather_chapter_5}
\begin{figure}
	\centerline{\includegraphics[width=0.45\textwidth]{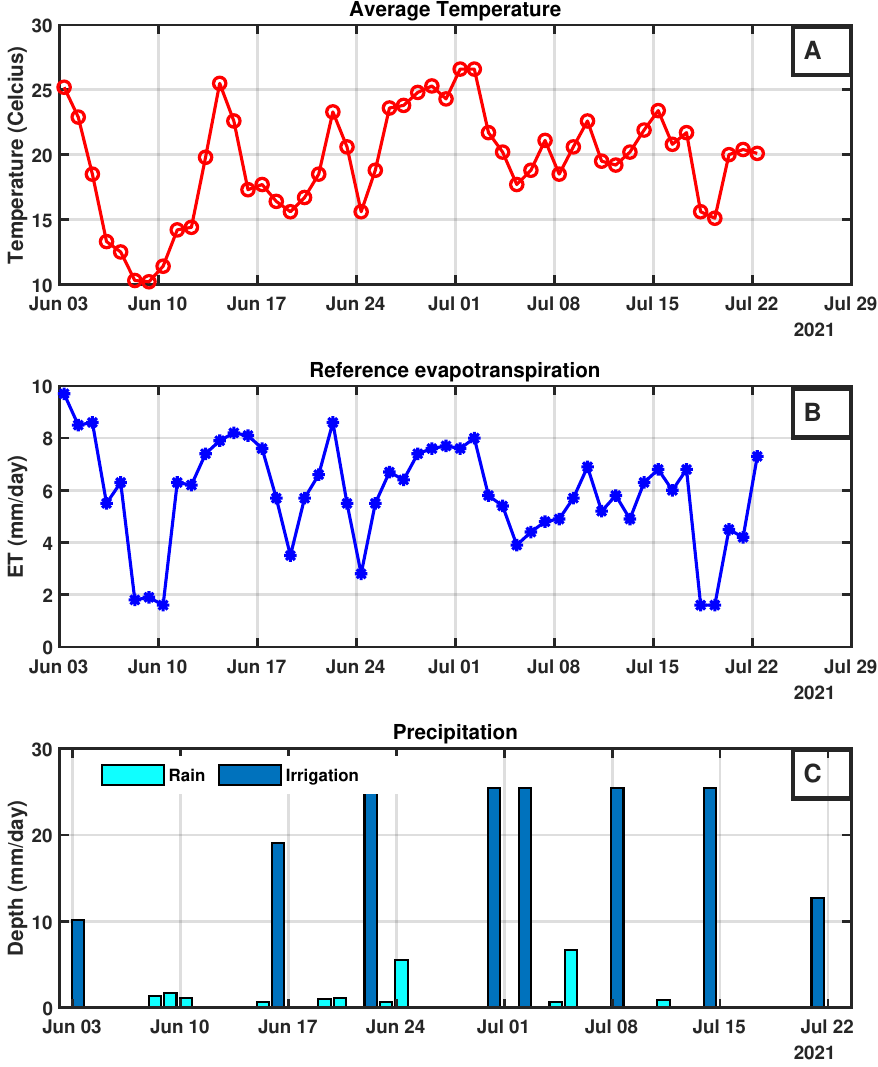}}
	\caption{Trajectories of the daily average temperature (\textbf{A}), daily reference evapotranspiration (\textbf{B}), and daily precipitation rates (\textbf{C}) for the entire simulation period.}
	\label{fig:et_rain_irrig_chapter_5}
\end{figure}

\end{document}